\documentclass[12pt]{article}
\usepackage{epsfig}
\usepackage{hhline}
\usepackage{times}
\usepackage{cite}

\newlength{\dinwidth}
\newlength{\dinmargin}
\newcommand{\pom}{{I\!\!P}}
\newcommand{\reg}{{I\!\!R}}

\newcommand {\lapprox}
   {\raisebox{-0.7ex}{$\stackrel {\textstyle<}{\sim}$}}
\def\gsim{\,\lower.25ex\hbox{$\scriptstyle\sim$}\kern-1.30ex%
\raise 0.55ex\hbox{$\scriptstyle >$}\,}
\def\lsim{\,\lower.25ex\hbox{$\scriptstyle\sim$}\kern-1.30ex%
\raise 0.55ex\hbox{$\scriptstyle <$}\,}

\newcommand{\PO}{I\!\!P}

\newcommand{\xpom}{x_{\PO}}

\def\Journal#1#2#3#4{{#1} {\bf #2} (#3) #4}

\def\NIMA{{\em Nucl. Instrum. Methods} {\bf A}}
\def\NPB{{\em Nucl. Phys.}   {\bf B}}

\def\PLB{{\em Phys. Lett.}   {\bf B}}
\def\PRL{\em Phys. Rev. Lett.}
\def\PRD{{\em Phys. Rev.}    {\bf D}}
\def\ZPC{{\em Z. Phys.}      {\bf C}}
\def\EJC{{\em Eur. Phys. J.} {\bf C}}
\def\CPC{\em Comp. Phys. Commun.}

\newcommand{\av}[1]{\mbox{$ \langle #1 \rangle $}}

\usepackage{dcolumn}
\newcolumntype{d}[1]{D{.}{.}{#1}}

\begin{document}  

\begin{titlepage}

\begin{flushleft}
DESY 00--174 \hfill ISSN 0418--9833 \\
December 2000
\end{flushleft}

\vspace{2.2cm}

\begin{center}
\begin{Large}

{\bf  Diffractive Jet Production \\ 
in Deep-Inelastic {\boldmath$ e^+p$} Collisions at HERA}

\vspace{1.8cm}

H1 Collaboration

\end{Large}
\end{center}

\vspace{1.3cm}

\begin{abstract}
\noindent
A measurement is presented of dijet and 3-jet cross sections in
low-$|t|$ diffractive deep-inelastic scattering interactions of the
type $ep \rightarrow eXY$, where the system $X$ is separated by a
large rapidity gap from a low-mass baryonic system $Y$.  Data taken
with the H1 detector at HERA, corresponding to an integrated
luminosity of $18.0 \rm\ pb^{-1}$, are used to measure hadron level
single and double differential cross sections for $4<Q^2<80 \rm\ 
GeV^2$, $x_\pom<0.05$ and $p_{T,jet}>4 \rm\ GeV$.  The energy flow not
attributed to jets is also investigated.  The measurements are
consistent with a factorising diffractive exchange with trajectory
intercept close to 1.2 and tightly constrain the dominating
diffractive gluon distribution.  Viewed in terms of the diffractive
scattering of partonic fluctuations of the photon, the data require
the dominance of $q\overline{q}g$ over $q\overline{q}$ states. Soft
colour neutralisation models in their present form cannot
simultaneously reproduce the shapes and the normalisations of the
differential cross sections. Models based on 2-gluon exchange are able
to reproduce the shapes of the cross sections at low $x_\pom$ values.
\end{abstract}

\vspace{1cm}

\begin{center}
submitted to \EJC
\end{center}


\end{titlepage}

\begin{flushleft}

C.~Adloff$^{33}$,              
V.~Andreev$^{24}$,             
B.~Andrieu$^{27}$,             
T.~Anthonis$^{4}$,             
V.~Arkadov$^{35}$,             
A.~Astvatsatourov$^{35}$,      
I.~Ayyaz$^{28}$,               
A.~Babaev$^{23}$,              
J.~B\"ahr$^{35}$,              
P.~Baranov$^{24}$
% ...

\bigskip{\it
% ...
$ ^{4}$ Inter-University Institute for High Energies ULB-VUB, Brussels;
          Universitaire Instelling Antwerpen, Wilrijk; Belgium$^{ c}$ \\
% ...
$ ^{23}$ Institute for Theoretical and Experimental Physics,
          Moscow, Russia \\
 $ ^{24}$ Lebedev Physical Institute, Moscow, Russia$^{ e,h}$ \\
% ...
$ ^{27}$ LPNHE, Ecole Polytechnique, IN2P3-CNRS, Palaiseau, France \\
 $ ^{28}$ LPNHE, Universit\'{e}s Paris VI and VII, IN2P3-CNRS,
          Paris, France \\
% ...
$ ^{33}$ Fachbereich Physik, Bergische Universit\"at Gesamthochschule
          Wuppertal, Wuppertal, Germany$^{ a}$ \\
% ...
$ ^{35}$ DESY, Zeuthen, Germany$^{ a}$ \\
% ...
$ ^a$ Supported by the Bundesministerium f\"ur Bildung, Wissenschaft,
      Forschung und Technologie, FRG,
      under contract numbers 7AC17P, 7AC47P, 7DO55P, 7HH17I, 7HH27P,
      7HD17P, 7HD27P, 7KI17I, 6MP17I and 7WT87P \\
% ...
$ ^c$ Supported by FNRS-NFWO, IISN-IIKW \\
% ...
$ ^e$ Supported by the Deutsche Forschungsgemeinschaft \\
% ...
$ ^h$ Supported by Russian Foundation for Basic Research
      grant no. 96-02-00019 \\
% ...
}

\end{flushleft}

\newpage
\setcounter{page}{1}


\section{Introduction}
\label{chapter0}

The observation of deep-inelastic scattering (DIS) events at HERA
containing a large gap in the rapidity distribution of the final state
hadrons \cite{obsdiff} has generated considerable renewed interest in
understanding colour singlet exchange in strong interactions.
At high energy, such interactions are interpreted as being due to
diffractive scattering.
HERA has made it possible to study diffraction
using a highly virtual photon probe. This
offers the chance to illuminate the underlying dynamics in terms of
quantum chromodynamics (QCD).

Inclusive diffractive DIS is principally sensitive
to the role of quarks in the scattering 
process \cite{H1:F2d93,H1:F2d94,f2dzeus}. 
More insight into the gluonic degrees of freedom can be obtained by
studying the hadronic final 
state \cite{H1:diffhfs,diffhfs-zeus,djzeus,H1:d2j94}. 
Final states containing heavy quarks or high
transverse momentum ($p_T$) jets
are of particular interest, 
since the additional hard scales may ensure the applicability
of perturbative QCD 
techniques \cite{gg1,gg,bartelsqq,bartelsqqg}.  
High $p_T$ jet production in diffraction has previously been studied both in 
$p\overline{p}$ collisions \cite{difjetsUA8,difjetsCDF,facbreak,difjetsD0} 
and at HERA \cite{djzeus,H1:d2j94}.

In this article, a high statistics measurement of diffractive jet 
production is presented, which was performed using the H1 detector.
The data were obtained using events 
where the proton (or a low-mass proton excitation) loses only
a small fraction of its incoming momentum and escapes undetected
through the beam pipe.  Separated from this system by a large rapidity
region devoid of  activity, the hadronic system $X$
is well contained within the central part of the detector and contains
the high $p_T$ jets.
The luminosity is increased by an order of magnitude compared with 
previous H1 measurements \cite{H1:d2j94}
and the kinematic range is also extended.
This makes it possible to extract double differential cross sections
for the first time and to study 3-jet as well as dijet production.

The dijet data yield direct constraints on the diffractive gluon distribution
and are used to investigate the QCD \cite{collins}
and Regge \cite{IngSchl} factorisation properties
of diffractive DIS. QCD inspired models \cite{bartelsqq,bartelsqqg,sat}
based on the exchange
of a pair of gluons from the proton \cite{lownussinov}
are compared with the data in a
restricted kinematic region where they are most likely to be applicable.
Predictions from soft colour neutralisation 
models \cite{semicl,sci,scinew}
are also
confronted with the data. 

The article is organised as follows. The kinematics of diffractive
scattering at HERA are introduced in section \ref{chapter1}. In
section \ref{chapter2}, an overview of phenomenological models and QCD
calculations relevant for diffractive jet production is given and the
Monte Carlo simulation of diffractive events is described. In section
\ref{chapter3}, the data selection, the cross section measurement
procedure and the determination of the systematic uncertainties are
explained. The results, expressed in terms of hadron level single and
double differential cross sections, are presented and discussed in
section \ref{chapter4}.  


\section{Diffractive Scattering at HERA}
\label{chapter1}

\subsection{Inclusive Diffractive Scattering}

\begin{figure}[t]
\centering \epsfig{file=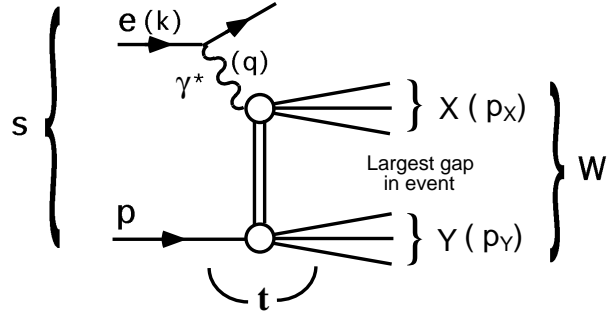,width=.5\linewidth}
\caption{ The generic diffractive process at HERA, where the 
  electron ($k$) couples to a photon ($q$) which interacts with the
  proton ($P$) via net colour singlet exchange, producing two distinct
  final state hadronic systems $X$ and $Y$.  If the masses of $X$ and
  $Y$ are small compared with $W$, the two systems are separated by a
  large gap in rapidity. }
\label{diagrams}
\end{figure}

Fig.~\ref{diagrams} illustrates the generic diffractive process at
HERA of the type $ep\rightarrow eXY$.  The electron\footnote{From now
  on, the word `electron' will be used as a generic term for electrons
  and positrons.}  (with 4-momentum $k$) couples to a virtual photon
$\gamma^*$ ($q$) which interacts with the proton ($P$). The usual DIS
kinematic variables are defined as
\begin{equation}
Q^2=-q^2 \ ; \qquad y=\frac{P\cdot q}{P \cdot k} \ ; \qquad 
x=\frac{-q^2}{2 P\cdot q} \ .
\end{equation}
The squared invariant masses of the electron-proton and photon-proton systems
$s$ and $W^2$ are given by
\begin{equation}
s=(k+P)^2\simeq{Q^2}/{xy} \simeq (300 \ \mathrm{GeV})^2 \ ; \qquad
 W^2=(q+P)^2 \simeq ys-Q^2 \ .
\label{eq:w2}
\end{equation}
If the interaction takes place via colour singlet exchange, the photon
and proton dissociate to produce distinct hadronic systems $X$ and
$Y$, with invariant masses $M_X$ and $M_Y$ respectively.  In the case
where $M_X$ and $M_Y$ are small compared with $W$, the two systems are
separated by a large rapidity gap. The longitudinal momentum fraction
$x_\pom$ of the colourless exchange with respect to the incoming
proton and the squared four-momentum transferred at the proton vertex
$t$ are then defined by
\begin{equation}
\xpom = \frac{q \cdot (P - p_Y)}{q \cdot P} \ ; \qquad 
t=(P-p_Y)^2 \ ,
\end{equation}
where $p_Y$ is the 4-momentum of $Y$. In the analysis presented here,
$t$ and $M_Y$ are not measured and hence are integrated over
implicitly\footnote{It is noted that for this analysis $M_Y=M_p$
  dominantly.}.  In addition, the quantity $\beta$ is defined as
\begin{equation}
  \beta = \frac{x}{x_\pom} = \frac{Q^2}{2 q \cdot (P - p_Y)} \ .
\end{equation}
In an interpretation in which partonic structure is ascribed to the
colourless exchange, $\beta$ is the longitudinal momentum fraction of
the exchange that is carried by the struck quark, in analogy to $x$ in
the case of inclusive scattering.

\subsection{Diffractive Dijet Production}
\label{sec:jetkin}

\begin{figure}[t]
\centering
\epsfig{file=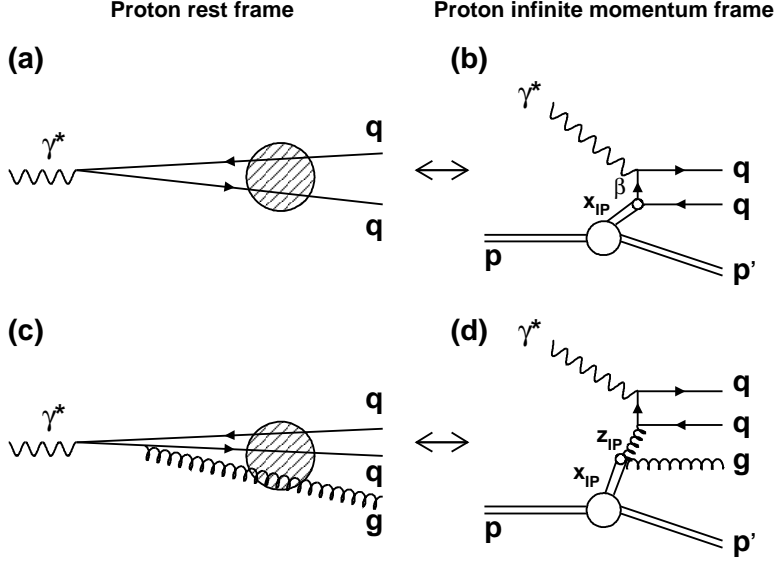,width=0.67\linewidth}
\caption{ 
  Diffractive scattering in the proton rest frame and the proton
  infinite momentum frame (figure after \cite{semicl}). In the proton
  rest frame, the virtual photon dissociates into a $q\overline{q}$
  state ({\em a}), scattering off the proton by colour singlet (e.g.
  2-gluon) exchange. In the infinite momentum frame, this can be
  related to diffractive quark scattering ({\em b}).  The emission of
  an additional gluon forms an incoming $q\overline{q}g$ state ({\em
    c}). If the gluon is the lowest $p_T$ parton, this contribution
  can be related to diffractive Boson-Gluon-Fusion ({\em d}).  }
\label{diagrams2}
\end{figure}

Viewing DIS at low $x$ in the proton rest frame, the virtual photon
splits into a $q\overline{q}$ pair well in advance of the 
proton target (fig.~\ref{diagrams2}a). The $q\overline{q}$ state may then
scatter elastically with the proton.  The production of high $p_T$
final states by the diffractive $q \overline{q}$ scattering process is
heavily suppressed \cite{AJM} and the invariant masses $M_X$ produced
are typically small.  It is thus expected that for large values of
$M_X$ or $p_T$, ${\cal O} (\alpha_s)$ contributions due to the
radiation of an extra gluon become important \cite{gg1,bmh}. The
result is an incoming $q\overline{q}g$ system (fig.~\ref{diagrams2}c).

In the proton infinite momentum frame, the lowest order (i.e. ${\cal
  O} (\alpha_s^0)$) contribution to the diffractive cross section is
the quark scattering diagram (fig.~\ref{diagrams2}b).  The
${\cal O} (\alpha_s)$ contributions are Boson-Gluon-Fusion
(BGF) and QCD-Compton (QCDC) scattering.  Unlike inclusive diffractive
scattering, jet production is directly sensitive to the
role of gluons in diffraction due to the direct coupling to the gluon
in the BGF diagram (fig.~\ref{diagrams2}d).

There is a correspondence between the proton rest frame and the
infinite momentum frame pictures, which is discussed here in the
context of the leading $\log (Q^2)$ approximation.  
For the dominant configuration
in which the photon longitudinal momentum is shared asymmetrically between
the partons,
diffractive
$q\overline{q}$ scattering (fig.~\ref{diagrams2}a) can be related to
the diffractive quark scattering diagram (fig.~\ref{diagrams2}b).  If
the gluon is the lowest $p_T$ parton, the diffractive scattering of
asymmetric $q\overline{q}g$ configurations
(fig.~\ref{diagrams2}c) can be related to diffractive BGF
(fig.~\ref{diagrams2}d).  If the $q$ or $\overline{q}$ is the lowest
$p_T$ parton, the process corresponds to diffractive QCDC scattering
(not shown).  

Using the non-zero invariant mass squared $\hat{s}$ of the two
highest $p_T$ partons emerging from the hard interaction in the ${\cal
  O} (\alpha_s)$ case, the quantity $z_\pom$ is introduced:
\begin{equation}
   z_\pom = \beta \cdot (1+\hat{s}/Q^2) \ .
\label{eq:zpomdef}
\end{equation}
Similarly to $\beta$ for the case of the lowest order diagram 
(fig.~\ref{diagrams2}b),
$z_\pom$ corresponds to the longitudinal momentum fraction of the
exchange which takes part in the hard interaction (fig.~\ref{diagrams2}d).


\section{Phenomenological Models and Monte Carlo Simulation}
\label{chapter2}

In this section, several phenomenological approaches and QCD
calculations are discussed, which attempt to describe diffractive DIS,
including diffractive jet production. The focus is on the models which
are compared with the data in section \ref{chapter4}.

\subsection{Diffractive Parton Distributions}
\label{section:difpdf}

In the leading $\log(Q^2)$ approximation, the cross section for the
diffractive process $\gamma^*p\rightarrow p'X$ can be written in terms
of convolutions of universal partonic cross sections
$\hat{\sigma}^{\gamma^*i}$ with diffractive parton distributions
$f_i^D$, representing probability distributions for a parton $i$ in
the proton under the constraint that the proton remains intact with
particular values of $x_\pom$ and $t$. Thus, at leading twist,
\begin{equation}
\frac{{\rm d^2} \sigma(x,Q^2,x_\pom,t)^{\gamma^*p\rightarrow p'X}}
{{\rm d} x_\pom \ {\rm d} t} \ = \ 
\sum_i \int_x^{x_\pom}{\rm d}\xi \
\hat{\sigma}^{\gamma^*i}(x,Q^2,\xi) \ 
f_i^D(\xi,Q^2,x_\pom,t) \ .
\label{equ:diffpdf}
\end{equation}
This factorisation formula holds for large enough $Q^2$ and fixed $x$,
$x_\pom$ and $t$.  This ansatz, introduced in \cite{trentadue,berera},
was rigorously proven for inclusive diffractive lepton-hadron
scattering in \cite{facproof,collins}.  The diffractive parton
distributions are not known from first principles, though they should
obey the DGLAP \cite{dglap} evolution equations.  Recently, there have
been attempts to calculate the diffractive parton distributions at a
starting scale $\mu_0^2$ for QCD evolution under certain assumptions.
In \cite{hautmann}, the proton is replaced by a small-size pair of
heavy quarks, such that perturbation theory can be applied.  A
different approach is the semiclassical model by Buchm\"uller,
Gehrmann and Hebecker \cite{semicl}, based on the opposite extreme of
a very large hadron. In spite of the different assumptions, the two
approaches give rather similar results for the diffractive parton
distributions. The general behaviour is the same as the momentum
fractions tend to 0 or 1 and the gluon distribution dominates.

\subsection{Resolved Pomeron Model and Pomeron Parton Distributions}
\label{chapter:rapgapmodel}

The application of Regge phenomenology of soft hadronic high energy
interactions to the concept of diffractive parton distributions
(section~\ref{section:difpdf}) leads to the Ingelman-Schlein model of a
`resolved pomeron' with a partonic structure \cite{IngSchl} invariant
under changes in $\xpom$ and $t$.  The diffractive parton
distributions then factorise into a flux factor $f_{\pom/p}$ and
pomeron parton distributions $f_i^{\pom}$:
\begin{equation}
f_i^D(x,Q^2,x_\pom,t) \ = \ f_{\pom /p}(x_\pom ,t)  \cdot 
 f_i^{\pom }(\beta=x/x_\pom ,Q^2) \ .
\label{reggefac}
\end{equation}
The universal flux factor describes the probability of finding a
pomeron in the proton as a function of $x_\pom$ and $t$. The pomeron
parton distributions are usually expressed in terms of
$\beta$.

The triple differential cross section for inclusive diffraction ${\rm
  d}^3 \sigma / {\rm d} \beta \, {\rm d} Q^2 \, {\rm d} \xpom$ is
often presented in the form of a diffractive structure function
$F_2^{D(3)}(\beta, Q^2, \xpom)$.  In \cite{H1:F2d94}, the H1
collaboration interpreted a measurement of $F_2^{D(3)}$ 
in terms of a resolved pomeron model: At the largest $x_\pom$
studied, it was necessary to consider more generally contributions
from sub-leading reggeon exchanges\footnote{Throughout this paper, the
  term `reggeon' ($\reg$) will be used to describe this contribution.}
as well as the pomeron, such that (neglecting possible interference
terms)
\begin{equation}
f_i^D (x, Q^2, \xpom, t) \ = \  
f_{\pom /p} (\xpom, t) \cdot f_i^\pom(\beta,Q^2)  +  
f_{\reg /p} (\xpom, t) \cdot f_i^\reg(\beta,Q^2) \ .
\label{h1flux}
\end{equation}
The flux factors for the pomeron and reggeon exchanges were
parameterised in a Regge-inspired form:
\begin{equation}
f_{\{\pom,\reg\}/p}(\xpom , t) \ = \ C_{\{\pom,\reg\}} \ 
{\xpom}^{1-2\alpha_{\{\pom,\reg\}}(t)} \ e^{b_{\{\pom,\reg\}} t} \ ,
\label{eq:flux}
\end{equation}
with
$\alpha_{\{\pom,\reg\}}(t)=\alpha_{\{\pom,\reg\}}(0)+\alpha'_{\{\pom,\reg\}}
t$.  From fits in which the parton densities evolve according to the
DGLAP equations, parameterisations of the pomeron quark and gluon
distributions and values for the trajectory intercepts 
$\alpha_\pom(0)$ and $\alpha_\reg(0)$
were obtained.  The resulting value of $\alpha_\pom (0)= 1.203 \pm
0.020 \ ({\rm stat.}) \pm 0.013 \ ({\rm syst.}) \pm 0.030 \ ({\rm
  model})$ is significantly higher than that obtained from soft
hadronic interactions, where $\alpha_\pom(0) \simeq 1.08$
\cite{DL:stot}. The parton densities extracted for the pomeron are
dominated by gluons, which carry $80 - 90\%$ of the exchanged momentum
throughout the measured $Q^2$ range.

\subsection{Colour Dipole and 2-Gluon Exchange Models}
\label{ggsec}

In the proton rest frame, diffractive DIS is often treated by
considering the $q \overline{q}$ and $q \overline{q} g$ photon fluctuations
(fig.~\ref{diagrams2}a,c) as (effective) colour dipoles. The
diffractive $\gamma^*p$ cross section can be factorised into a 
squared
effective photon dipole wave function and a squared `dipole cross
section' for the scattering of these dipoles off the proton
\cite{nikzak90,mcdermott00}.  The gross features of the diffractive $\beta$
distribution can be deduced from a knowledge of the partonic wave
functions of the photon alone.  According to a recent QCD motivated
parameterisation \cite{bekw}, longitudinally and transversely polarised
$q\overline{q}$ states dominate at high and medium values of $\beta$
respectively, whereas the $q\overline{q}g$ state originating from
transversely polarised photons is dominant at low $\beta$.

Investigating diffractive final states with varying $p_T$ probes the
dipole cross section as a function of the dipole size.  Large size,
low $p_T$ configurations interact with the proton similarly to soft
hadron-hadron scattering.  Small size, high $p_T$ dipole
configurations lead to hard scales which encourage a perturbative QCD
treatment of the dipole cross section.  The precise dynamics of the
dipole cross section are not known a priori. However, the simplest
realisation of a net colour singlet exchange at the parton level is a
pair of gluons with cancelling colour charges \cite{lownussinov}.  We
focus below on two recent colour dipole models
\cite{sat,bartelsqq,bartelsqqg} based on 2-gluon exchange, where the
cross section is related to the square of the unintegrated gluon
distribution of the proton $\mathcal{F}(x,k_T^2)$ \cite{bfkl}.  Here,
$k_T$ is the parton transverse momentum relative to the proton
direction.  Other colour dipole approaches can be found in
\cite{gg1,gg,mcdermott}.

The dipole approach has been employed in the `saturation' model by
Golec-Biernat and W\"usthoff \cite{sat}. Here, an ansatz for the
dipole cross section is made which interpolates between the
perturbative and non-perturbative regions of $\sigma^{\gamma^*p}$.
This model is able to give a reasonable description of $F_2(x,Q^2)$ at
low $x$, which determines the three free parameters of the model.  The
parameterised dipole cross section can be re-expressed in terms of
$\mathcal{F}(x,k_T^2)$, such that the diffractive cross section is
predicted at $t = 0$.  Introducing an additional free parameter $B=6.0
\rm\ GeV^{-2}$ to describe the $t$ dependence as $e^{B t}$, the
diffractive structure function $F_2^{D(3)}$ is successfully
described.  The calculation of the $q\overline{q}g$ cross section is
made under the assumption of strong $k_T$ ordering of the final state
partons (leading $\log (Q^2)$ approximation), corresponding to
$k^{(g)}_T \ll k^{(q,\overline{q})}_T$.

Cross sections for diffractive $q\overline{q}$ and $q\overline{q}g$
production by 2-gluon exchange have been calculated by Bartels, Ewerz,
Lotter and W\"usthoff ($q\overline{q}$) \cite{bartelsqq} and by
Bartels, Jung, Kyrieleis and W\"usthoff ($q\overline{q}g$)
\cite{bartelsqqg}.  The derivative of the next-to-leading order (NLO)
GRV gluon parameterisation \cite{GRVfgluon} is used for
$\mathcal{F}(x,k_T^2)$.  The calculation of the $q\overline{q}g$ final
state is performed in the leading $\log (1/\beta)$, leading $\log
(1/x_\pom)$ approximation, such that configurations without strong
$k_T$ ordering are included. The calculations require all outgoing
partons to have high $p_T$ and are thus not suited to describe
$F_2^{D(3)}$.  The minimum value $p^{\rm cut}_{T,g}$ for the final 
state gluon
transverse momentum is a free parameter which can be used to tune
the model to the overall dijet cross section.  
As for the saturation model,
the calculation yields predictions at $t = 0$. The extension to finite
$t$ is performed using the Donnachie-Landshoff elastic proton form
factor \cite{dl:form}.  The sum of the $q \overline{q}$ and $q
\overline{q} g$ contributions in this model is hereafter referred to
as `BJLW'.

\subsection{Soft Colour Neutralisation Models}
\label{scisec}

An alternative approach to diffractive DIS is given by soft colour
neutralisation models, which naturally lead to very similar properties
of inclusive and diffractive DIS final states. In the Soft Colour
Interaction (SCI) model by Edin, Ingelman and Rathsman \cite{sci}, the
hard interaction in diffractive DIS is treated identically to that in
inclusive DIS.  Diffraction occurs through soft colour rearrangements
between the outgoing partons, leaving their momentum configuration
unchanged. If two colour singlet systems are produced by such a
mechanism, the hadronic final state can exhibit a large rapidity gap.
In the original SCI model, diffractive final states are produced using
only one free parameter, the universal colour rearrangement
probability, which is fixed by a fit to $F_2^{D(3)}$.  The model has
been refined recently \cite{scinew} by making the colour rearrangement
probability proportional to the normalised difference in the
generalised areas of the string configurations before and after the
rearrangement.

The semiclassical model, which was already
mentioned in section \ref{section:difpdf},
is a non-perturbative model based on the
dipole approach. Viewed in the proton rest
frame, $q\overline{q}$ and $q\overline{q}g$ photon fluctuations
scatter off a superposition of soft colour fields associated with the
proton. Those configurations which emerge in a net colour singlet
state contribute to the diffractive cross section \cite{bmh}. Assuming
a specific model for the proton wave functional \cite{semicl}, the
results are formulated as a parameterisation of $t$-integrated
diffractive parton distributions \cite{hebecker}, which are determined
from a combined four parameter fit to $F_2$ and $F_2^{D(3)}$ at low $x$ and
$x_\pom$.

\subsection{Monte Carlo Simulation}
\label{section:mc}

Monte Carlo simulations are used to determine the corrections to be
applied to the data to compensate for the limited efficiencies,
acceptances and resolutions of the detector.  The generated Monte
Carlo events are passed through a detailed simulation of the H1
detector and are subjected to the same reconstruction and analysis
chain as the data.

The main Monte Carlo generator used to correct the data is RAPGAP
2.08/06 \cite{rapgap}. Events are generated according to a resolved
(partonic) pomeron model (section~\ref{chapter:rapgapmodel}).
Contributions from pomeron and reggeon exchanges are included
neglecting any possible interference effects.  The parameterisations
of the pomeron and reggeon flux factors and parton distributions are
taken from the H1 analysis of $F_2^{D(3)}$ \cite{H1:F2d94}.  The
pomeron and reggeon trajectories and slope parameters
(eq.~\ref{eq:flux}) are $\alpha_\pom(t)=1.20+0.26t$, $b_\pom=4.6 \ 
\mathrm{GeV}^{-2}$, $\alpha_\reg(t)=0.50+0.90t$ and $b_\reg=2.0 \ 
\mathrm{GeV}^{-2}$.  The pomeron parton distributions are taken from 
the `flat
gluon' (or `fit 2' of \cite{H1:F2d94}) solution in 
the leading order DGLAP fits to $F_2^{D(3)}$. Those of the meson are
taken from fits to pion data \cite{owens}. The renormalisation and
factorisation scales are set to $\mu^2=Q^2+p_T^2$, where $p_T$ is the
transverse momentum of the partons emerging from the hard scattering
relative to the collision axis in the $\gamma^*p$ centre-of-mass
frame\footnote{This frame is also called the `hadronic centre-of-mass
  frame'.}.  The parton distributions are convoluted with hard
scattering matrix elements to $\mathcal{O}(\alpha_s)$.  Intrinsic
transverse momentum of the partons in the pomeron \cite{nikolaev} is
not included.  Charm quarks are produced in the massive scheme via
Boson-Gluon-Fusion. For the production of light quarks, a lower
cut-off in $p_T^2$ is introduced in the ${\cal O} (\alpha_s)$ QCD
matrix elements to avoid divergences.  Higher order QCD diagrams are
approximated with parton showers in the leading $\log(Q^2)$
approximation (MEPS) \cite{meps} or the colour dipole
approach\footnote{ `Colour dipole approach' as an approximation to
  higher order QCD effects should not be confused with the `Colour
  dipole models' introduced in section \ref{ggsec}.}  (CDM) \cite{cdm} as
implemented in ARIADNE \cite{ariadne}.  Hadronisation is simulated
using the Lund string model in JETSET \cite{lund}.  QED radiative
effects are taken into account via an interface to the HERACLES
program \cite{heracles}.

The RAPGAP simulation includes 
a contribution of events where the virtual photon
$\gamma^*$ is assigned an internal partonic structure.
The resolved virtual photon is parameterised according to
the SaS-2D \cite{sas} set of photon parton densities, which has been
found to give a reasonable description of inclusive dijet production
at low $Q^2$ \cite{h1:virtgam}.

Monte Carlo generators are also used to compare the measured hadron
level cross sections with the predictions of the phenomenological
models and QCD calculations described in sections 
\ref{chapter:rapgapmodel}-\ref{scisec}.  All
of the predictions are made to leading order of QCD. Unless otherwise
stated, higher order QCD effects are approximated by initial and final
state parton showers.  RAPGAP is used to obtain the predictions of the
resolved pomeron model with different pomeron intercept values and
parton distributions.  It also contains implementations of the
saturation, semiclassical and BJLW models.  Both versions of the SCI
model are implemented in the LEPTO $6.5.2\beta$ generator
\cite{lepto}.


\section{Experimental Procedure}
\label{chapter3}

The analysis presented in this article is based on H1 data taken in
the years 1996 and 1997, when HERA collided $E_e=27.5 \rm\ GeV$
positrons with protons of $E_p=820 \ \mathrm{GeV}$.  The data
correspond to an integrated luminosity of $18.0 \rm\ pb^{-1}$.  A
detailed description of the measurement can be found in
\cite{fpschill}.  This section begins with a short overview of the H1
detector, after which the data selection is described. Then, the cross
section measurement and the determination of the systematic errors are
explained.

\subsection{H1 Detector}
\label{detector:section}

The H1 detector is described in detail elsewhere \cite{H1:det}.  Here,
we give a brief description of the detector components most relevant
for the present analysis. The $z$ axis of the H1 coordinate system
corresponds to the beam axis such that positive $z$ values
refer to the direction of the outgoing proton beam, often called the
`forward' direction\footnote{This direction corresponds to positive
  values of the pseudorapidity $\eta=-\ln\tan\theta/2$.}.

The interaction region is surrounded by the tracking system. Two large
concentric drift chambers (CJC), located within a solenoidal magnetic
field of $1.15 \rm\ T$, measure the trajectories of charged particles
and hence their momenta in the range $-1.5<\eta<1.5$. The resolution
is $\sigma(p_T)/p_T \simeq 0.01 p_T/\mathrm{GeV}$.  Energies of final
state particles are measured in a highly segmented Liquid Argon (LAr)
calorimeter covering the range $-1.5<\eta<3.4$, surrounding the
tracking detectors. The energy resolution is $\sigma(E)/E \simeq
11\%/\sqrt{E/\mathrm{GeV}}$ for electromagnetic showers and
$\sigma(E)/E \simeq 50\%/\sqrt{E/\mathrm{GeV}}$ for hadrons, as
obtained from test beam measurements.  The overall hadronic energy
scale of the LAr is known to $4\%$.  The backward direction
($-4.0<\eta<-1.4$) is covered by a lead / scintillating fibre
calorimeter (SPACAL) with electromagnetic and hadronic sections. The
energy resolution for electrons is $\sigma(E)/E \simeq
10\%/\sqrt{E/\mathrm{GeV}}$.  The energy scale uncertainty is $0.3\%$
for electrons with $E'_e=27.5 \rm\ GeV$ and $2.0\%$ at $E'_e=8 \rm\ 
GeV$.  The electron polar angle is measured to $1 \rm\ mrad$.  The
energy scale of the SPACAL is known to $7\%$ for hadrons.  In
front of the SPACAL, the Backward Drift Chamber (BDC) provides track
segments of charged particles with a resolution of $\sigma (r) =0.4
\rm\ mm$ and $r \sigma (\phi) =0.8 \rm\ mm$.  The $ep$ luminosity is
determined with a precision of $2\%$ by comparing the measured event
rate in a photon tagger calorimeter close to the beam pipe at $z=-103
\ \mathrm{m}$ with the QED Bremsstrahlung ($ep\rightarrow ep\gamma$)
cross section.

To enhance the sensitivity to hadronic activity in the region of the
outgoing proton, the Forward Muon Detector (FMD) and the Proton
Remnant Tagger (PRT) are used.  The FMD is located at $z=6.5 \ 
\mathrm{m}$ and covers the pseudorapidity range $1.9<\eta<3.7$
directly. Particles produced at larger $\eta$ can also be detected
because of secondary scattering with the beam-pipe. The PRT, a set of
scintillators surrounding the beam pipe at $z=26 \ \mathrm{m}$, can
tag hadrons in the region $6.0 \ \lapprox \ \eta \ \lapprox \ 7.5$.

\subsection{Data Selection}
\label{sec:datasel}

DIS events are triggered by an electromagnetic energy cluster in the
SPACAL in coincidence with a CJC track. Scattered electron candidates are
then selected with $E'_e>8 \ \mathrm{GeV}$ in the angular range
$156^\circ<\theta'_e<176^\circ$. Various cuts are applied on these
candidates in order to select electrons and reject background from
photons and hadrons. Among these are requirements on the width of the
shower, the containment within the electromagnetic part of the SPACAL
and the existence of an associated track segment in the BDC.  The $z$
coordinate of the reconstructed vertex is required to lie within
$\pm35 \ \mathrm{cm}$ ($\pm \sim \! \! 3\sigma$) of the nominal
interaction point. To suppress events with initial state QED
radiation, the summed $E-p_z$ of all reconstructed final state
particles including the electron\footnote{For events fully contained
  in the detector, the total $E-p_z$ is sharply peaked at $2 E_e = 55
  \rm\ GeV$.} has to be greater than $35 \ \mathrm{GeV}$.  The DIS
kinematic variables are calculated from the polar angle and energy
measurements of the scattered electron:
\begin{equation}
Q^2=4E_eE'_e\cos^2\frac{\theta'_e}{2} \ ; \qquad 
y=1-\frac{E'_e}{E_e}\sin^2\frac{\theta'_e}{2} \ .
\end{equation}
Events which fulfil
$4<Q^2<80 \ \mathrm{GeV}^2$ and $0.1<y<0.7$ 
are selected.

The selection of diffractive events is based on the absence of
hadronic activity in the outgoing proton region. No signal above noise
levels is allowed in the FMD and PRT detectors. The most forward part
($\eta>3.2$) of the LAr calorimeter has to be devoid of hadronic
clusters with energies $E>400 \ \mathrm{MeV}$. This selection ensures
that the photon dissociation system $X$ is well contained within the
central part of the H1 detector and is separated by a large rapidity
gap covering at least $3.2 < \eta \ \lapprox \ 7.5$ from the $Y$
system.  The upper limit in $\eta$ implies that the $Y$ system escapes
undetected through the beam pipe and imposes the approximate
constraint $M_Y<1.6 \ \mathrm{GeV}$ and $|t|<1.0 \ \mathrm{GeV}^2$.

The $X$ system is measured in the LAr and SPACAL calorimeters together
with the CJC. It is reconstructed using a method that combines
calorimeter clusters and tracks whilst avoiding double counting
\cite{fscomb}. The dissociation mass is then calculated according to
\begin{equation}
M_X^2=(\textstyle\sum_{i} \displaystyle E_i)^2 - (\textstyle \sum_{i} 
\displaystyle {\bf p}_i)^2 \ ,
\end{equation}
where the sum runs over all reconstructed objects except for the
scattered electron\footnote{When calculating all hadronic final state
  quantities, particle masses are neglected.}.  $W^2$ is calculated
according to eq.~(\ref{eq:w2}).  $x_\pom$ and $\beta$ are then
computed from
\begin{equation}
x_\pom=\frac{Q^2+M_X^2}{Q^2+W^2} \ ; \qquad \beta=\frac{Q^2}{Q^2+M_X^2} \ .
\end{equation}
A cut $x_\pom<0.05$ is applied to suppress contributions from
non-diffractive scattering and secondary exchanges.  The resolution in
$\log x_\pom$ is approximately $8\%$.

The 4-vectors of the hadronic final state particles associated to the
$X$ system are Lorentz-transformed to the $\gamma^*p$ centre-of-mass
frame, where they are subjected to the CDF cone jet algorithm
\cite{cdfcone} with a cone radius of
$\sqrt{(\Delta\eta)^2+(\Delta\phi)^2}=1.0$.  The jets are required to
lie within the region $-1.0<\eta^{lab}_{jet}<2.2$ to ensure good
containment within the LAr calorimeter. Transverse energies and
momenta are calculated with respect to the $\gamma^*p$ axis.  Events
with either at least two or exactly three jets with transverse
momentum $p^*_{T,jet}>4 \ \mathrm{GeV}$ are selected for the dijet and
3-jet samples respectively. The average resolution in $p^*_{T,jet}$ is
$14\%$.  No requirements are made on the presence or absence of
hadronic activity beyond the jets.
The final event selection yields 2506 dijet and 130 3-jet events.

\begin{figure}[t]
\centering
\epsfig{file=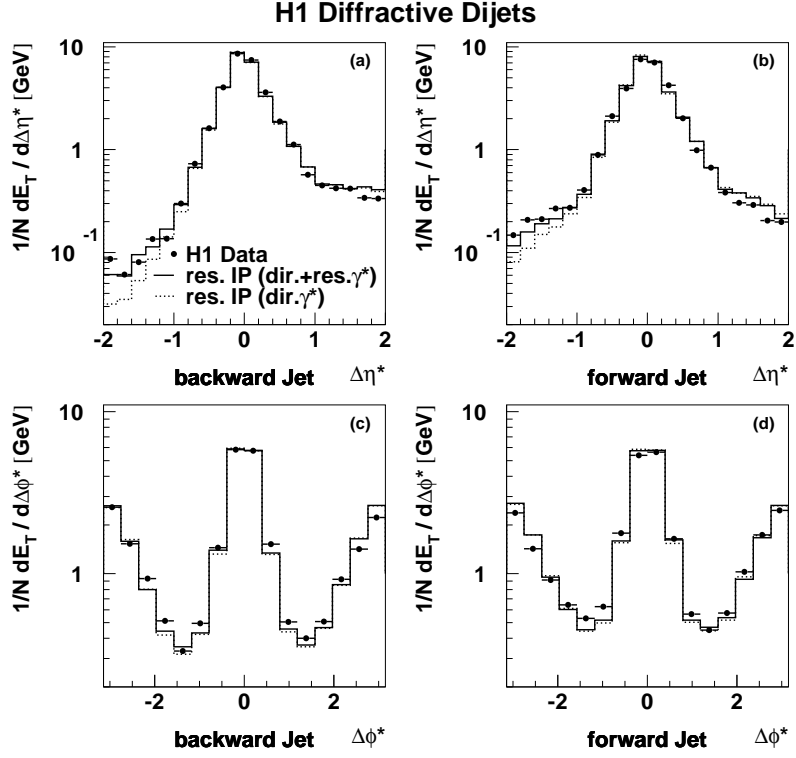,width=0.67\linewidth}
\caption{ Observed distributions of the average transverse energy flow
  per event around the jet axes in the diffractive dijet sample.
  $\Delta\eta^*$ and $\Delta\phi^*$ are the distances from the jet
  axes in pseudorapidity and azimuthal angle in the hadronic
  centre-of-mass frame.  The jet profiles in $\eta$ and $\phi$ are
  integrated over $\pm 1$ unit in $\phi$ and $\eta$ respectively.
  {\em(a)} and {\em(c)} show the distributions for the backward jet in
  the laboratory frame, whereas {\em(b)} and {\em(d)} show those for
  the forward jet. The distributions for the simulated sample of
  RAPGAP events are compared with the data.  Here, the contributions
  from direct photons only {\em (dotted histograms)} and from the sum
  of direct and resolved photon contributions {\em (solid histograms)}
  are shown.}
\label{fig9}
\end{figure}

Fig.~\ref{fig9} shows the transverse energy flow around the jet axes
for the dijet sample. For the jet profiles in $\eta$ and $\phi$, only
transverse energies within one unit in azimuth and pseudorapidity are
included in the plots respectively.  The jet profiles for backward and
forward jets are shown separately in Figs.~\ref{fig9}a,c and b,d.  The
data exhibit a clear back-to-back dijet structure in azimuth. The
energy flow is well described by the RAPGAP simulation that is used to
correct the data (solid lines).

\subsection{Cross Section Measurement}

The data are first corrected for losses at the trigger level, which
occur due to the track requirement. For the selected events, the
trigger efficiency varies between 80 and $90\%$, depending on the
kinematics.  Corrections for detector acceptances and migrations
between measurement intervals are evaluated by applying a bin-to-bin
correction method using the RAPGAP program (see section
\ref{section:mc}).  The simulation gives an acceptable description of
all relevant kinematic distributions of the selected dijet and 3-jet
events.  Smearing in $x_\pom$ is taken into account up to $x_\pom=0.2$
by the simulation of colour-singlet exchange in RAPGAP.  Migrations
from $\xpom > 0.2$ or from large values of $M_Y>5 \ \mathrm{GeV}$ are
covered by a RAPGAP simulation of inclusive DIS. This contribution is
at the level of $5\%$ averaged over all measured bins and is
concentrated at large $x_\pom$. An additional correction of
$-6.5\%\pm6.5\%$ is applied to account for the net smearing about the
$M_Y=1.6 \rm\ GeV$ boundary. Since only elastically scattered protons
have been simulated in RAPGAP, this correction is evaluated using the
proton dissociation simulation in the DIFFVM \cite{diffvm} Monte Carlo
model.  A further correction of $+5.5\%\pm1.4\%$ takes into account
diffractive events rejected due to fluctuations in the noise level in
the FMD. This correction is determined using randomly triggered
events.  QED radiative corrections are of the order of $5\%$. The 
bin purities and stabilities\footnote{`Bin purity' is defined as the
  fraction of simulated events reconstructed in a bin that are also
  generated in that bin. `Stability' is the fraction of events
  generated in a bin that are also reconstructed in that bin.} are
typically of the order of $50$ to $60\%$ and it is ensured that they
exceed $30\%$ for every measured data point.

The corrected cross sections are defined in a model independent
manner, whereby the systems $X$ and $Y$ are separated by the largest
gap in rapidity among the hadrons in the $\gamma^* p$ centre-of-mass
frame (fig.~\ref{diagrams}).  The $ep$ cross sections are corrected to
the hadron level and are quoted at the Born level.  The kinematic
range in which the cross sections are measured is fully specified in
tab.~\ref{kinrange}.  The measured range of jet pseudorapidities in
the hadronic centre-of-mass frame $-3<\eta_{jet}^*<0$ approximately
matches the $-1<\eta^{lab}_{jet}<2.2$ cut for the selected events.  No
$\eta_{\rm\max}$ or similar cuts are imposed in the definition of the
measured cross sections.

\begin{table}[t]
\begin{center}
\begin{tabular}{|c|} \hline
Kinematic Range of \\
Hadron Level Cross Sections \\ \hline  \hline
$4 < Q^2 < 80 \ {\rm GeV^2}$ \\ 
$0.1 < y < 0.7$ \\ \hline 
$\xpom < 0.05$ \\ 
$M_Y < 1.6 \ {\rm GeV}$ \\ 
$|t| < 1.0 \ {\rm GeV^2}$ \\ \hline
$N_{\rm jets} \geq 2 \ \ \mathrm{or} \ \  N_{\rm jets} = 3 $ \\ 
$p^*_{T,jet} > 4 \ {\rm GeV}$  \\
$-3 < \eta^*_{jet} < 0 $ \\ \hline
\end{tabular}
\end{center}
\caption{The kinematic range to which the dijet and 3-jet hadron level 
cross sections are corrected. The details of the jet finding 
algorithm can be found in section \ref{sec:datasel}.}
\label{kinrange}
\end{table}

\subsection{Analysis of Systematic Uncertainties}

The following sources of uncertainty contribute to the 
systematic errors on the measured cross sections.  The uncertainties
associated with detector understanding (see
section~\ref{detector:section}) are as follows.

\begin{enumerate}
  
\item The uncertainties in the hadronic calibrations of the LAr and
  SPACAL calorimeters mainly influence the measured values of
  $p^*_{T,jet}$ and $M_X$. The resulting uncertainties in the cross
  sections are up to $10\%$ (with a mean value of $5\%$) for the LAr
  and $0.5\%$ for the SPACAL.
  
\item The uncertainties in the $E'_e$ and $\theta'_e$ measurements
  propagate into the reconstruction of $Q^2$, $y$ and $W$ and the
  definition of the $\gamma^*p$ axis for the boost into the
  $\gamma^*p$ centre-of-mass frame.  The uncertainty in $\theta'_e$ leads to a
  systematic error of $1\%$ to $2\%$. The uncertainty in $E'_e$
  results in a systematic error between $1\%$ and $5\%$, depending on
  the kinematics.
  
\item The uncertainty in the fraction of energy of the reconstructed
  hadronic objects carried by tracks is $3\%$, leading to a systematic
  error in the range $1\%$ to $5\%$.
  
\item The uncertainties in the determinations of the trigger
  efficiency and the $ep$ luminosity affect the total normalisation by
  $5\%$ and $2\%$ respectively.
  
\item There is an uncertainty of $25\%$ in the fraction of events lost
  due to noise in the FMD, which translates into a $1.4\%$
  normalisation error on the measured cross sections.

\end{enumerate}

\noindent
The Monte Carlo modelling of the data gives rise to the following
uncertainties.

\begin{enumerate}
\setcounter{enumi}{5}

\item The uncertainty in the number of events migrating into the
  sample from $x_\pom>0.2$ or $M_Y>5 \rm\ GeV$ is estimated as $25\%$,
  leading to a systematic error between $1\%$ and $3\%$, with the
  biggest values at large $x_\pom$.
  
\item A $6.5\%$ uncertainty arises from the 
  correction for smearing about the $M_Y$ limit of the measurement.
  It is estimated by variations of: (a) the ratio of elastic proton to
  proton dissociation cross sections in DIFFVM between 1:2 and 2:1;
  (b) the generated $M_Y$ distribution within $1/M_Y^{2.0\pm0.3}$; (c)
  the $t$ dependencies in the proton dissociation simulation by
  changing the slope parameter by $\pm1 \rm\ GeV^{-2}$ and (d) the
  simulated efficiencies of the forward detectors FMD and PRT by
  $\pm4\%$ and $\pm25\%$ respectively.
  
\item The uncertainty arising from the QED radiative corrections is
  typically $5\%$, originating from the limited statistics of the
  Monte Carlo event samples.
  
\item The use of different approximations for higher order QCD
  diagrams (the parton shower (MEPS) model or the colour dipole (CDM)
  approach) leads to a $3\%$ uncertainty in the cross sections.
  
\item The model dependence of the acceptance and migration corrections
  obtained from the RAPGAP simulation is estimated by varying the
  shapes of kinematic distributions in the simulations beyond the
  limits imposed by previous measurements or the present data. This is
  done by reweighting (a) the $z_\pom$ distribution by
  $z_\pom^{\pm0.2}$ and $(1-z_\pom)^{\pm0.2}$; (b) the $p_T$
  distribution by $(1/p_T)^{\pm0.5}$; (c) the $x_\pom$ distribution by
  $(1/x_\pom)^{\pm0.2}$; (d) the $t$ distribution by $e^{\pm2t}$ and
  (e) the $\eta^{lab}_{jet}$ distribution to that observed in the
  data.  The resulting systematic uncertainties range between $6\%$
  and $13\%$, the largest contributions originating from (c) and (e).
  
\item The lower $p_T^2$-cut-off chosen to avoid collinear divergences
  in the leading order QCD matrix elements in RAPGAP is relatively
  high ($p_T^2>9 \rm\ GeV^2$) with respect to the experimental cut of
  $p^{* 2}_{T,jet}>16 \rm\ GeV^2$. Studying the dependence on the
  cut-off value results in an additional uncertainty of $5\%$.
\end{enumerate}

Most of the systematic uncertainties are not strongly correlated
between data points.  The total systematic error has been evaluated
for each data point by adding all individual systematic errors in
quadrature. It ranges between $15$ and $30\%$ and for most data points
is significantly larger than the statistical uncertainty.


\section{Results}
\label{chapter4}

In this section, the measured hadron level differential cross sections are 
presented for the kinematic region specified in
tab.~\ref{kinrange}. The cross sections are shown graphically in
Figs.~\ref{fig1a}-\ref{fig8}.  In all figures, the inner error bars
correspond to the statistical error, whilst the outer error bars
represent the quadratic sum of the statistical and systematic errors.
The numerical values of the measured cross sections can be found in
Tabs.~\ref{tab:xsa}-\ref{tab:xsg}.  The quoted differential cross
sections are average values over the intervals specified in the
tables.

\subsection{General Properties of the Dijet Data}
\label{generalprop}

\begin{figure}[t]
\centering
\epsfig{file=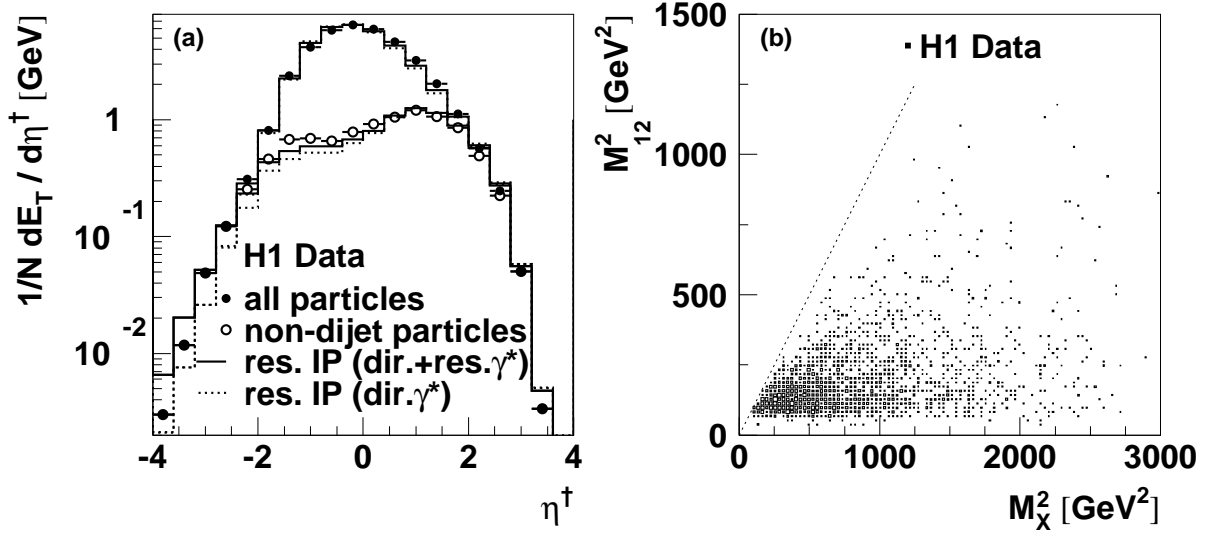,width=1.0\linewidth}
\caption{ {\em(a)} The uncorrected distribution of the average transverse 
  energy per event for the diffractive dijet sample
  as a function of the pseudorapidity $\eta^\dagger$
  in the centre-of-mass frame of the $X$ system.  
  Distributions are shown both for all final state
  particles {\em(solid points)} and for only those particles which do
  not belong to the two highest $p_T$ jets {\em (open points)}. The
  prediction of the RAPGAP simulations for direct and for direct plus
  resolved virtual photon contributions are also shown.  {\em(b)} The
  uncorrected correlation between the squared invariant mass of the
  $X$ system $M_X^2$ and the squared dijet invariant mass $M^2_{12}$
  for the diffractive dijet sample. The dotted line
  corresponds to $M_X^2=M_{12}^2$. }
\label{fig10}
\end{figure}

\begin{figure}[p]
\centering
\epsfig{file=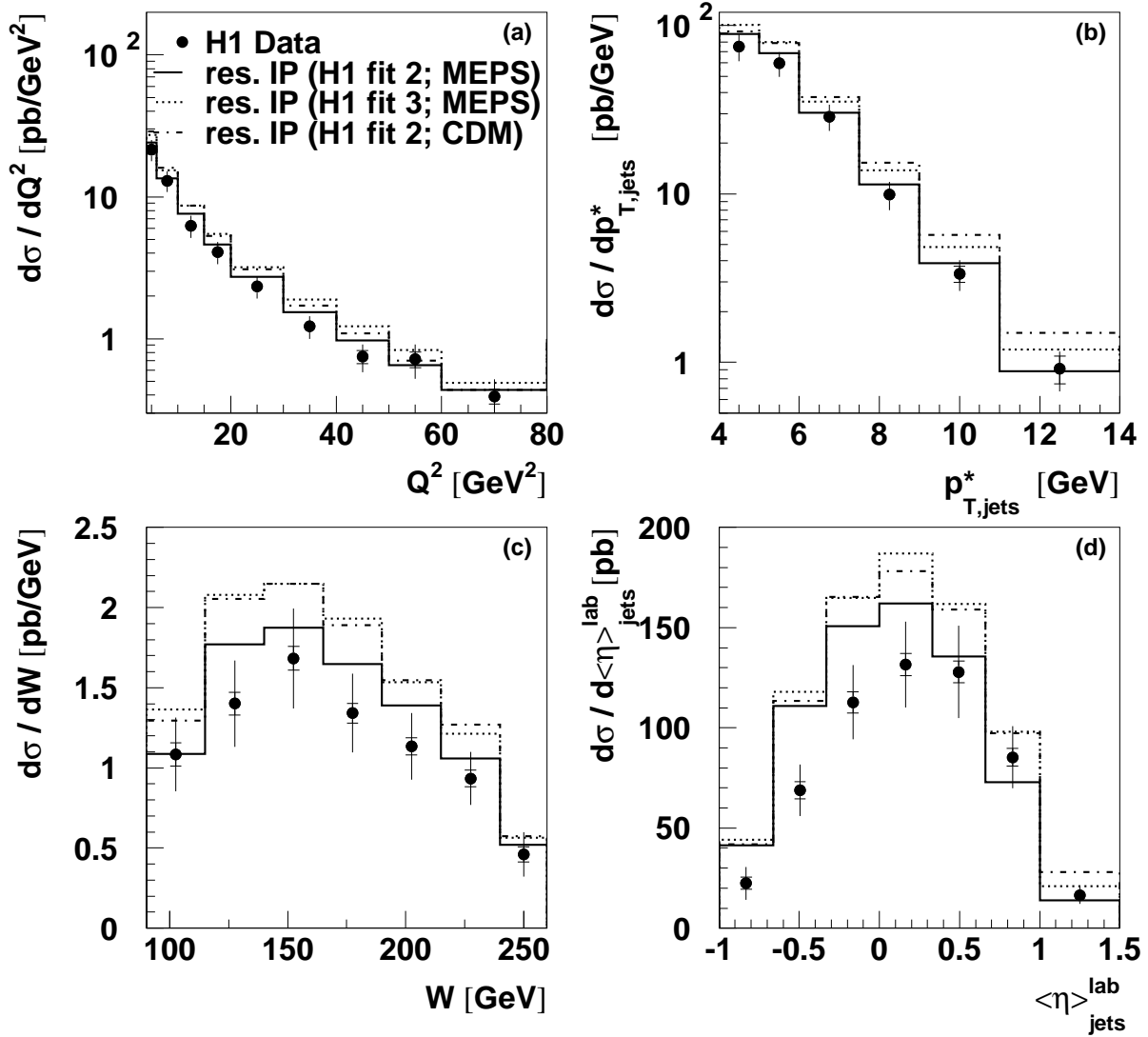,width=1.0\linewidth}
\caption{  Diffractive dijet cross sections as a function of {\em(a)} the
  photon virtuality $Q^2$, {\em(b)} the mean transverse jet momentum
  $p^*_{T,jets}$ in the $\gamma^* p$ centre-of-mass frame, 
  {\em(c)} the $\gamma^*p$ invariant mass $W$ and
  {\em(d)} the mean jet pseudorapidity in the laboratory frame
  $\av{\eta}^{lab}_{jets}$.  Also shown are the predictions from a
  resolved (partonic) pomeron model with gluon dominated pomeron
  parton distributions as obtained from the QCD analysis of
  $F_2^{D(3)}$ by H1 \cite{H1:F2d94}.  The results, using both the
  `fit 2' (`flat gluon') and `fit 3' (`peaked gluon') parton
  distributions for the pomeron, are shown evolved to a scale
  $\mu^2=Q^2+p_T^2$.  Resolved virtual photon contributions are added
  according to the SaS-2D parameterisation \cite{sas}. The prediction
  based on `fit 2' is also shown where the colour dipole approach (CDM)
   for higher order QCD effects is used in place of parton showers (MEPS). }
\label{fig1a}
\end{figure}

\begin{figure}[tb]
\centering
\epsfig{file=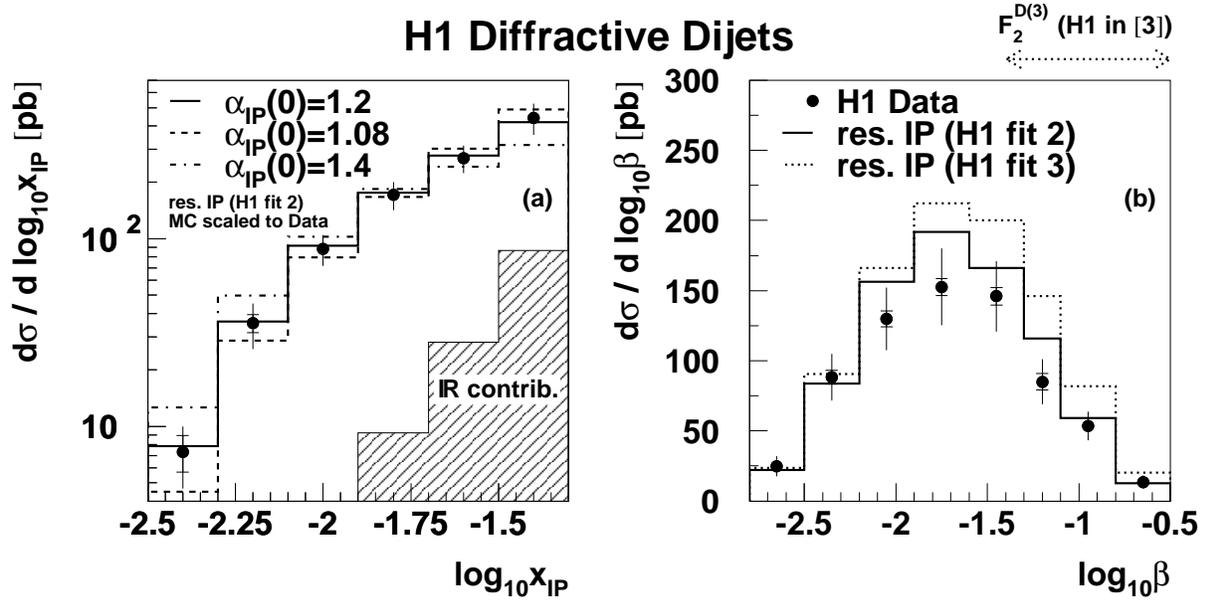,width=1.0\linewidth}
\caption{ Differential diffractive dijet cross sections as  functions of
  {\em(a)} $\log x_\pom$ and {\em(b)} $\log \beta$.
  The solid curves represent the
  predictions of the resolved pomeron model (`fit 2') as described in
  the text with direct and resolved photon contributions.  For the
  $\log x_\pom$ distribution, the contribution from sub-leading
  reggeon exchange is indicated by the
  hatched area. The dashed and dashed-dotted histograms correspond to
  the cross section predictions where the value of the pomeron
  intercept $\alpha_\pom(0)$ in the model was changed from the default
  value of 1.20 to 1.08 and 1.40 respectively. For this figure, all
  model predictions have been scaled to the integrated cross section in
the data.  For the $\log \beta$
  distribution, the prediction using the `fit 3' parton distributions
  is also shown and the range covered by the inclusive H1 measurement
  of $F_2^{D(3)}$ is indicated. }
\label{fig1b}
\end{figure}

\begin{figure}[tb]
\centering 
\epsfig{file=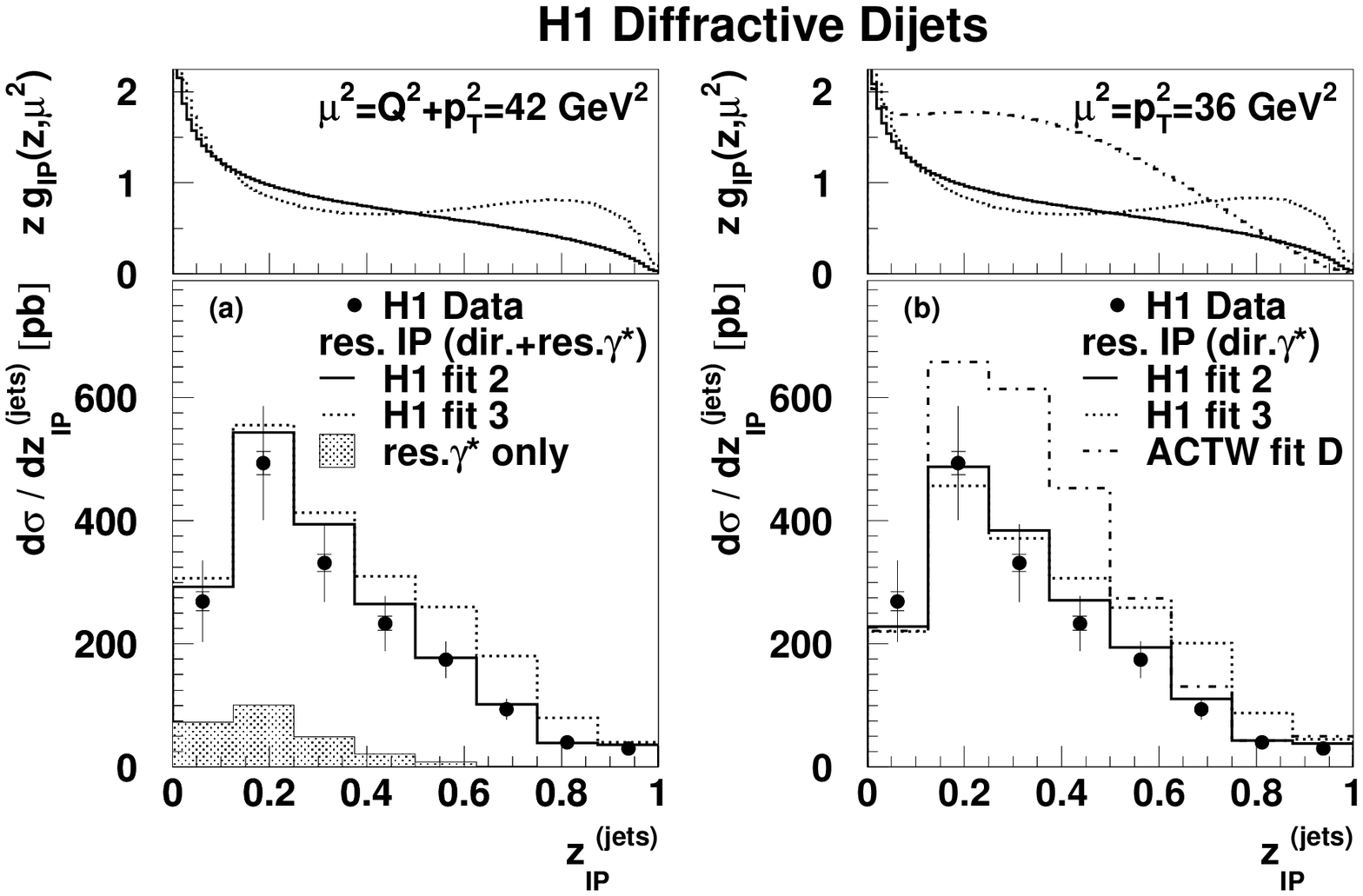,width=1.0\linewidth}
\caption{ The diffractive dijet cross section as a function of 
  $z^{(jets)}_\pom$.  The same data are compared to predictions of
  resolved pomeron models, where either {\em (a)} $\mu^2=Q^2+p_T^2$ or
  {\em (b)} $\mu^2=p_T^2$ are used as renormalisation and
  factorisation scales. In {\em (a)}, the `fit 2' (or `flat gluon')
  and `fit 3' (or `peaked gluon') parameterisations based on the H1
  leading order QCD fits to $F_2^{D(3)}$ \cite{H1:F2d94} are shown.
  Direct and resolved $\gamma^*$ contributions are both
  included. The size of the resolved $\gamma^*$ contribution in `fit
  2' is indicated by the shaded histogram.  In {\em (b)}, where only
  the direct $\gamma^*$ contributions are shown, the preferred
  solution `ACTW fit D' of the fits from \cite{actw} is shown in
  addition to the H1 fits.  The corresponding gluon distributions,
  evolved to the mean value of the respective scale used and
  normalised such that the pomeron flux $f_{\pom/p}(x_\pom=0.003,t=0)$
  is unity, are shown above the predictions. }
\label{fig3}
\end{figure}

In this section, general features of the data are discussed, referring
to Figs.~\ref{fig10}-\ref{fig3}. The model predictions\footnote{
  Software to produce predictions for the measured cross sections
  using any hadron level $ep$ Monte Carlo model is available in the
  HZTOOL package\cite{hztool}.}  which are also shown in these figures
are discussed in sections \ref{section:resultsrespom} and
\ref{section:xgam}.

In fig.~\ref{fig10}a, the uncorrected average transverse energy flow
per event for the dijet sample is shown   as a function of the
pseudorapidity $\eta^\dagger$ in the rest frame of the $X$
system\footnote{This frame can be interpreted as the $\gamma^* \pom$
  centre-of-mass frame. In this context, `$\pom$' or `pomeron' is used
  synonymously with `colourless exchange'.}.  
Positive values of $\eta^\dagger$
correspond to the pomeron hemisphere, negative values to the photon
hemisphere.  Both the total energy flow and the energy flow from
particles outside the two leading jets are shown.  The data exhibit
considerable hadronic energy not associated with the jets.  This
additional energy is distributed in both hemispheres with some
preference for the pomeron hemisphere.  In order to examine the
sharing of energy within the $X$ system on an event-by-event basis,
fig.~\ref{fig10}b shows the uncorrected correlation between the
squared dijet invariant mass $M_{12}^2$ and the squared total
diffractive mass $M_X^2$ \cite{bmh}.  $M_{12}$ is calculated from the
massless jet 4-vectors.  Except for a small subset of the events at
low $M_X$, only a fraction of the available energy of the $X$ system
is contained in the dijet system, such that $M_{12}$ is considerably
smaller than $M_X$ on average.

Figs.~\ref{fig1a} and \ref{fig1b} present differential dijet cross
sections as functions of the following observables: the photon
virtuality $Q^2$; the mean dijet transverse momentum $p_{T,jets}^*$,
defined as
\begin{equation}
p_{T,jets}^* = \textstyle \frac{1}{2}\displaystyle \left 
(p_{T,jet \, 1}^*+p_{T,jet \, 2}^*\right) \ ;
\end{equation}
the $\gamma^*p$ invariant mass $W$; the mean dijet pseudorapidity in the
laboratory frame $\av{\eta}^{lab}_{jets}$, defined as
\begin{equation}
\av{\eta}^{lab}_{jets} = \textstyle \frac{1}{2}\displaystyle 
\left(\eta_{jet \, 1}^{lab}+\eta_{jet \, 2}^{lab}\right) \ ;
\end{equation}
and the logarithms of the $x_\pom$ and $\beta$ variables.  The $Q^2$
and $p_{T,jets}^*$ distributions are steeply falling.  Due to the
selection of events with $Q^2>4 \ \mathrm{GeV}^2$ and $p_{T,jets}^{* \
  2}>16 \ \mathrm{GeV}^2$, the relation $p_{T,jets}^{* \ 2}>Q^2$ holds 
for the bulk
of the data. As can be seen in fig.~\ref{fig1a}c, the $W$ range of the
selected events is approximately $90<W<260 \rm\ GeV$.  The $x_\pom$
distribution shows a rising behaviour from the lowest accessible
values of $\sim 0.003$ up to the cut value of $0.05$. For kinematic
reasons, the dijet measurement is dominated by larger $x_\pom$ values
than is the case for inclusive diffractive measurements.  The $\beta$
range covered by the measurement extends down to almost $10^{-3}$,
lower than accessed so far in measurements of $F_2^{D(3)}$.  The
shapes of the measured cross sections are generally well described by
the RAPGAP simulation used to correct the data (solid histograms),
except for the $\av{\eta}^{lab}_{jets}$ distribution, which shows that
on average the measured jets have slightly larger pseudorapidities
than is predicted by the simulations.

In fig.~\ref{fig3}, the cross section is shown differentially in
$z^{(jets)}_\pom$, which is calculated from
\begin{equation}
z_\pom^{(jets)}=\frac{Q^2+M^2_{12}}{Q^2+M_X^2} \ .
\end{equation}
Monte Carlo studies show that the resolution in $z_\pom^{(jets)}$ is
approximately $25\%$ and that there is a good correlation between
$z_\pom^{(jets)}$ and the value of $z_\pom$ as defined at the parton
level in eq.~\ref{eq:zpomdef}.  In loose terms, the $z_\pom^{(jets)}$
observable measures the fraction of the hadronic final state energy of
the $X$ system which is contained in the two jets.  The measured
$z_\pom^{(jets)}$ distribution is largest around 0.2 and thus confirms
the observation from fig.~\ref{fig10} that the total energy of the $X$
system is typically much larger than that contained in the jets.
Diffractively scattered $q \overline{q}$ photon fluctuations (see
section~\ref{sec:jetkin}) satisfy $z_\pom \equiv 1$ at the parton
level, but can be smeared to $z_\pom^{(jets)}$ values as low as 0.6
because of fragmentation and jet resolution effects.  Even taking this
smearing into account, the $z_\pom^{(jets)}$ distribution implies the
dominance of $q\overline{q}g$ over $q\overline{q}$ scattering in the
proton rest frame picture.


\subsection{Interpretation within a Partonic Pomeron Model}
\label{section:resultsrespom}

In this section, the data are compared with the Ingelman-Schlein model
(section \ref{chapter:rapgapmodel}), using the
RAPGAP Monte Carlo model with various sets of pomeron parton
distributions.  In all cases unless otherwise stated, the RAPGAP
predictions shown use the parton shower approximation to higher order
diagrams (MEPS) and a contribution from resolved virtual photons is
included, as described in section \ref{section:mc}.  It has been shown
in an H1 measurement of inclusive dijet production for similar ranges
in $Q^2$ and $p^*_{T,jets}$ \cite{h1:virtgam} that including 
resolved photon contributions improves
the description of the data by leading order Monte Carlo Models in the
region $p_{T,jets}^{* \ 2}>Q^2$.  It is thus reasonable to expect a similar
contribution in diffraction.

\subsubsection{Diffractive Gluon Distribution}

Pomeron parton densities dominated by gluons have proved successful in
describing not only inclusive measurements of the diffractive
structure function \cite{H1:F2d93,H1:F2d94,f2dzeus}, but also more
exclusive hadronic final state analyses
\cite{djzeus,H1:d2j94,H1:diffhfs,diffhfs-zeus}.  By contrast, pomeron
parton distributions dominated by quarks (e.g. `fit 1' from
\cite{H1:F2d94}) do not describe the data
\cite{H1:F2d94,H1:d2j94,H1:diffhfs}. In particular, they lead to
significantly smaller predicted dijet electroproduction cross sections
than were obtained in previous measurements \cite{H1:d2j94}.  The
free parameters of the Ingelman-Schlein model to which dijet
production is most sensitive are the pomeron gluon distribution
$g_\pom(z,\mu^2)$ and the pomeron intercept $\alpha_\pom(0)$.  The
sub-leading reggeon contribution and the pomeron quark distribution
are better constrained by inclusive colour singlet exchange 
measurements \cite{H1:F2d94,baryons}.

Predictions based on two sets of pomeron parton distributions obtained
from the leading order DGLAP analysis of $F_2^{D(3)}$ from H1 in
\cite{H1:F2d94} are compared with the data in
Figs.~\ref{fig1a},~\ref{fig1b}.  The `flat gluon' or `fit 2'
parameterisation gives a very good description of all differential
distributions, except for ${\rm d}\sigma/{\rm
  d}\av{\eta}_{jets}^{lab}$.  
The predictions based on the
`peaked gluon' or `fit 3' parameterisation in
Figs.~\ref{fig1a},~\ref{fig1b} are also in fair agreement with the
data, though the description is somewhat poorer than that from `fit
2'.  If the colour dipole approximation (CDM) to higher order QCD
effects is used instead of parton showers (MEPS), the predicted dijet
cross sections increase in normalisation by approximately $15\%$
(fig.~\ref{fig1a}). The shapes of the predicted distributions,
including that of $z_\pom^{(jets)}$, are not significantly affected.

The cross section differential in $z^{(jets)}_\pom$ (fig.~\ref{fig3})
is also compared with predictions from different sets of pomeron
parton distributions.  Fig.~\ref{fig3}a shows the predictions based on
the partons extracted in `fit 2' and `fit 3' of \cite{H1:F2d94}.  The
parton distributions are evaluated at a scale\footnote{Alternative
  reasonable choices of scale such as $Q^2+4p_T^2$ make only small
  differences to the Monte Carlo predictions.}  $\mu^2=Q^2+p_T^2$.
The contribution of quark induced processes in the predictions is
small.  The fraction of the cross section ascribed to resolved virtual
photons, which is shown separately for `fit 2' in fig.~\ref{fig3}a, is
also small and is concentrated at low $z^{(jets)}_\pom$. The same is
true for the reggeon contribution (not shown).  
The predictions based on the `flat
gluon' or `fit 2' parton densities are in very good agreement with the
data.  The `peaked gluon' or `fit 3' parameterisation leads to an
overestimate of the dijet cross section at high values of
$z^{(jets)}_\pom$.  The gluon distributions from which the predictions
are derived are shown above the data at $\mu^2= 42 \ \mathrm{GeV}^2$,
representing the mean value of $Q^2+p_{T,jets}^{* \ 2}$ 
for the selected events.
The difference in shape between the gluon distributions and the hadron
level predictions reflects the kinematic range of the measurement
(tab.~\ref{kinrange}).  The dijet data are highly sensitive to the
shape of the gluon distribution, which 
is poorly constrained by the
$F_2^{D(3)}$ measurements. This is
especially the case in the region of large
momentum fractions ($z_\pom$ or $\beta$), 
since data with $\beta>0.65$ were excluded
from the DGLAP analysis of $F_2^{D(3)}$.

In fig.~\ref{fig3}b, the same data are compared with the models where
$p_T^2$ was chosen as the renormalisation and factorisation scale and
only direct photon contributions are included.  The level of agreement
between the data and the simulations based on the H1 fits is similar
to that in fig.~\ref{fig3}a. Also shown is a prediction based on the
best combined fit in \cite{actw} to H1 and ZEUS $F_2^{D(3)}$ data and
ZEUS diffractive dijet photoproduction data\footnote{In this
  parameterisation, the pomeron intercept is set to
  $\alpha_\pom(0)=1.19$.}.  Due to the different shape and
normalisation of the gluon distribution in this parameterisation, the
agreement with the dijet data is significantly poorer than is the case
for the two H1 fits.

In general, the close agreement between the `fit 2' and `fit 3'
parameterisations and the data can be interpreted as support for 
factorisable pomeron parton distributions in DIS, strongly dominated
by gluons with a momentum distribution relatively flat in $z_\pom$.

\begin{figure}[tb]
\centering
\epsfig{file=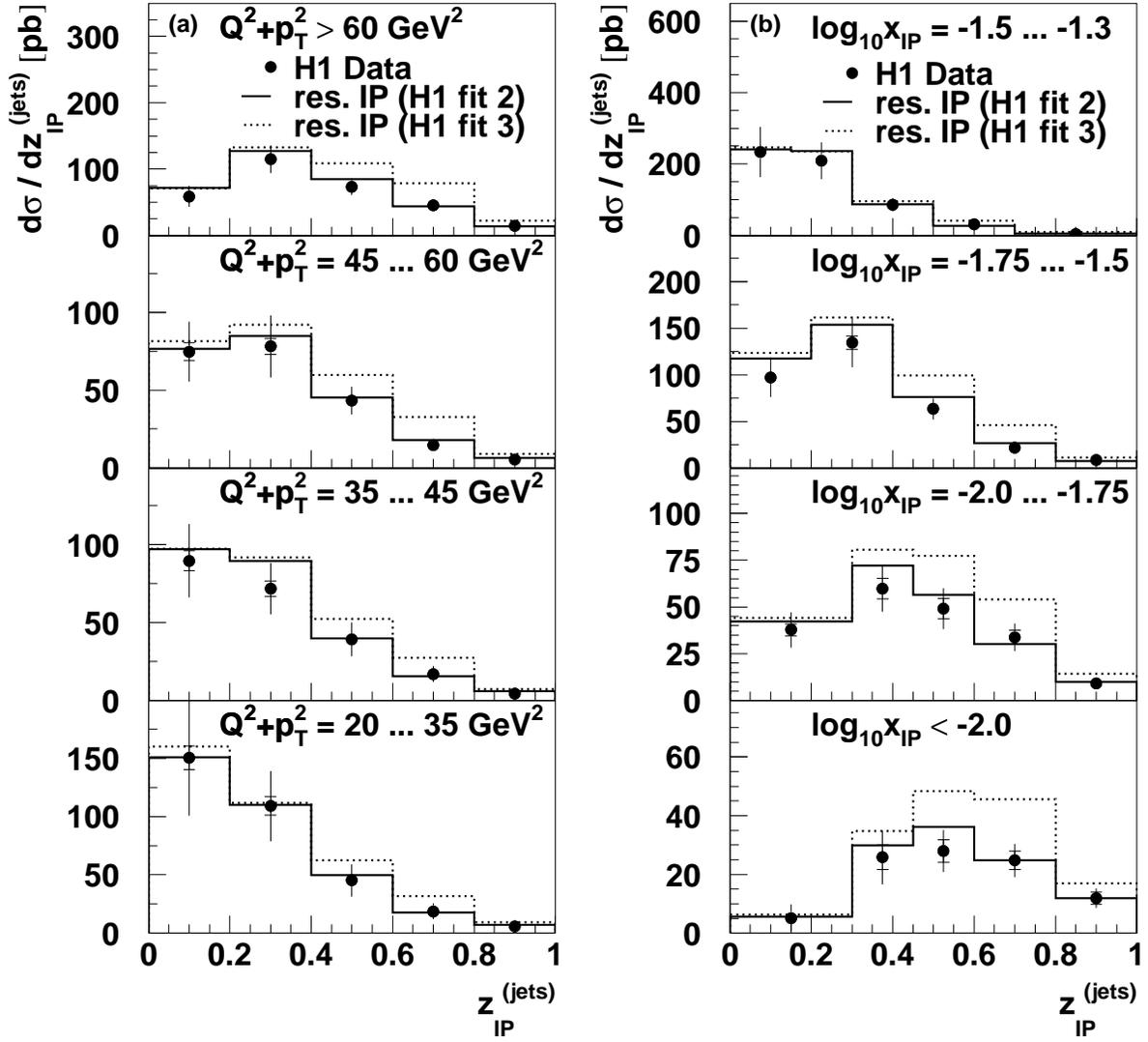,width=1.0\linewidth}
\caption{  Diffractive dijet cross sections as a function of
  $z^{(jets)}_\pom$, shown in four intervals of {\em (a)} the scale
  $\mu^2=Q^2+p_T^2$ and {\em (b)} $\log x_\pom$.  The data are
  compared to the resolved pomeron model based on the two fits to
  $F_2^{D(3)}$ from H1, including both direct and resolved $\gamma^*$
  contributions.  }
\label{fig4}
\end{figure}

\subsubsection{Scale Dependence, Regge Factorisation and Pomeron Intercept}

In the following, some basic assumptions of the resolved pomeron model
are tested, namely the evolution of the parton distributions with
scale, Regge factorisation and the universality of the pomeron
intercept.

Fig.~\ref{fig4}a shows the cross section differential in
$z^{(jets)}_\pom$ in four intervals of the scale $\mu^2=Q^2+p_T^2$.
Even in this double differential view, the `fit 2' resolved pomeron
model with parton densities evolving according to the DGLAP equations
gives a very good description of the data.  The `peaked gluon'
solution overestimates the cross section at high $z^{(jets)}_\pom$ in
all regions of $\mu^2$.

In fig.~\ref{fig4}b, the data are used to test Regge factorisation
(eq.~\ref{reggefac}).  The cross
section differential in $z^{(jets)}_\pom$ is measured in four
intervals of $x_\pom$.  A substantial dependence of the shape of the
$z^{(jets)}_\pom$ distribution on $x_\pom$ is observed. This is
dominantly a kinematic effect, since $x_\pom$ and $z^{(jets)}_\pom$
are connected via the relation $x_\pom \cdot z_\pom^{(jets)} =
x_p^{(jets)}$, where $x_p^{(jets)}$ is the proton momentum fraction
which enters the hard process.  The range in $x_p^{(jets)}$ is
approximately fixed by the kinematic range of the measurement.  Again,
the factorising resolved pomeron model describes the distributions
well.  Thus, at the present level of precision, the data are
compatible with Regge factorisation.  There is little freedom to
change the pomeron intercept $\alpha_\pom(0)$ and compensate this by
adjusting the gluon distribution. Fast variations of $\alpha_\pom(0)$
with $z_\pom$ are also incompatible with the data.

The value of $\alpha_\pom(0)$ controls the energy or $x_\pom$
dependence of the cross section. In the predictions of the resolved
pomeron model shown in Figs.~\ref{fig1a}-\ref{fig4}, a value of
$\alpha_\pom(0)=1.2$ is used, as obtained in the H1 analysis of
$F_2^{D(3)}$ \cite{H1:F2d94}.  Since this value of $\alpha_\pom(0)$ is
larger than that describing soft interactions, it is interesting to
investigate whether further variation takes place with the additional
hard scale introduced in the dijet sample. In fig.~\ref{fig1b}a, the
effect on the shape of the predicted cross section differential in
$x_\pom$ is investigated when $\alpha_\pom(0)$ is varied. As examples,
the predictions with $\alpha_\pom(0) = 1.08$ (`soft pomeron') and
$\alpha_\pom(0) = 1.4$ (approximate leading order `BFKL pomeron'
\cite{bfkl}) are shown.  All predictions have been scaled to the 
total cross section in the data.
The $\xpom$ dependence of the data requires a value for
$\alpha_\pom(0)$ close to 1.2. The values of 1.08 and 1.4 result in
$x_\pom$ dependences which are steeper or flatter than the data
respectively.  Making a fit for $\alpha_\pom(0)$ to the shape of the
$x_\pom$ cross section, assuming a flux of the form given in 
eq.~\ref{eq:flux}, yields a value of
\begin{eqnarray}
\alpha_\pom(0) =  1.17 \ \pm 0.03 \ ({\rm stat.}) \ 
                         \pm 0.06 \ ({\rm syst.}) \ 
                         ^{+0.03}_{-0.07} \ ({\rm model}) \ . \nonumber 
\end{eqnarray}
The model dependence uncertainty is evaluated by varying the resolved
photon and the reggeon contributions in the model by $\pm 50 \%$ each,
changing the pomeron gluon distribution within the range allowed by
the measured $z_\pom^{(jets)}$ distribution, varying the assumed
$\alpha'_\pom$ within $0.26 \pm 0.26 \rm\ GeV^{-2}$ and varying
$b_\pom$ between $2 \rm\ GeV^{-2}$ and $8 \rm\ GeV^{-2}$.
The effects of NLO corrections and possible
pomeron-reggeon interference have not been studied.
The extracted value of $\alpha_\pom(0)$ is compatible with that obtained 
from inclusive diffraction in a similar $Q^2$ region, despite the
fact that the jets introduce an additional hard scale.

\subsection{Energy Flow in the Photon Hemisphere and Resolved 
Virtual Photons} 
\label{section:xgam}

As can be seen from Figs.~\ref{fig1a}-\ref{fig4}, the data are well
described by the resolved pomeron model, where a contribution from
resolved virtual photons is included as described in
section~\ref{section:mc}.  In this section, two observables are
studied which are particularly suited to the interpretation of the
data in terms of direct and resolved photon contributions.

\begin{figure}[tb]
\centering
\epsfig{file=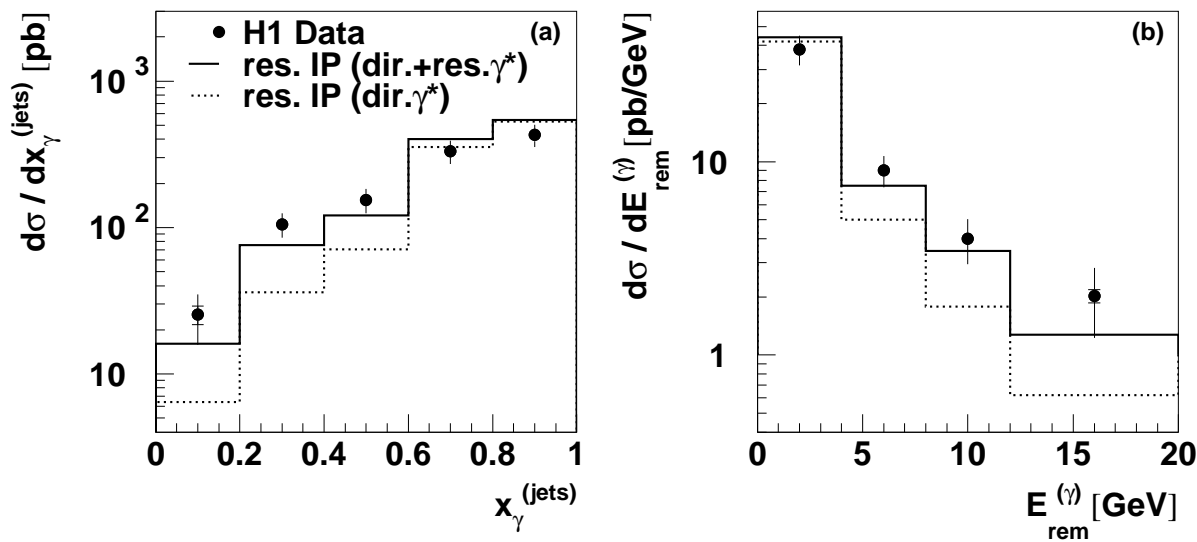,width=1.0\linewidth}
\caption{  Differential diffractive dijet cross sections as a function of 
  {\em (a)} $x_\gamma^{(jets)}$, an estimator for the photon momentum
  fraction entering the hard scattering process, and {\em (b)}
  $E_{rem}^{(\gamma)}$, the summed hadronic final state energy not
  belonging to the two highest $p^*_T$ jets in the photon hemisphere of
  the $\gamma^*\pom$ centre-of-mass frame.  The data are compared to
  the resolved pomeron model (`fit 2') with and without an additional
  contribution from resolved virtual photons, parameterised according
  to the SaS-2D photon parton distributions.}
\label{fig2}
\end{figure}

As in the case of real photoproduction analyses (see e.g.
\cite{gammap}), a quantity $x_\gamma$ is defined as the fraction of
the photon momentum which enters the hard scattering. If the 4-vector
of the parton from the photon entering the hard scattering is labelled
$u$, then
\begin{equation}
x_\gamma=\frac{P\cdot u}{P\cdot q} \ .
\end{equation}
Direct photon events satisfy $x_\gamma \equiv 1$ by definition. Events where
the photon is resolved have $x_\gamma<1$. At the hadron level, an
observable $x_\gamma^{(jets)}$ can be constructed by measuring the
ratio of the summed $E-p_z$ of the two jets to the total $E-p_z$:
\begin{equation}
x_\gamma^{(jets)}=\frac{\sum_{jets} E-p_z}{\sum_{all} E-p_z} \ .
\end{equation}
The observable $x_\gamma^{(jets)}$ 
correlates well with the parton level $x_\gamma$
and is reconstructed with a resolution of approximately $12\%$
relative to the hadron level definition.  The cross section
differential in $x_\gamma^{(jets)}$ is shown in fig.~\ref{fig2}a. The
distribution is peaked at values around 1 but there is also a sizeable
cross section at lower $x_\gamma^{(jets)}$ values. The prediction of
the resolved pomeron model with only direct photon contributions
describes the high $x_\gamma^{(jets)}$ region, but lies significantly
below the data at low values of $x_\gamma^{(jets)}$. The prediction is
non-zero in this region only because of migrations from the parton level value
of $x_\gamma$ to the hadron level quantity $x_\gamma^{(jets)}$. If the
contribution from resolved photons is included, a much improved
description of the data is achieved.  The total predicted dijet cross
section then increases by $17\%$.

The part of the hadronic final state not associated to the two highest
$p^*_T$ jets is best studied in the $\gamma^*\pom$ centre-of-mass frame
(see section~\ref{generalprop}).  Hadronic final state particle
production outside the two highest $p^*_T$ jets can originate from
jet resolution effects, 
possible photon and pomeron remnants or from higher order QCD
diagrams.  In order to further investigate the energy in the photon
hemisphere, a new observable $E_{rem}^{(\gamma)}$ is constructed.
$E_{rem}^{(\gamma)}$ is defined as the energy sum of all final state
hadrons in the photon hemisphere ($\eta^\dagger <0$) which lie outside
the two highest $p^*_T$ jet cones.  The cross section is shown
differentially in $E_{rem}^{(\gamma)}$ in fig.~\ref{fig2}b.  The
distribution falls quickly as $E_{rem}^{(\gamma)}$ increases,
indicating the dominance of
direct photon scattering. The description at high
$E_{rem}^{(\gamma)}$ values (corresponding to $x_\gamma<1$) is again
much improved by adding the resolved $\gamma^*$ contribution.

The presence of resolved virtual photon contributions is also
suggested by the energy flow backward of the jets (corresponding to
the photon direction) in the jet profiles (fig.~\ref{fig9}).
Similarly, the transverse energy not associated with the jets in the
$\eta^\dagger < 0$ hemisphere of the $\gamma^*\pom$ system
(fig.~\ref{fig10}a), is best described when the resolved photon
contribution is added.  Good descriptions of these distributions
cannot be achieved by adjusting the diffractive gluon distribution.
The resolved virtual photon contributions can be viewed as an
approximation to NLO QCD diagrams and/or contributions without strong
$k_T$ ordering.  The possible presence of such effects will be
investigated further in section \ref{2gluon}.

\subsection{Soft Colour Neutralisation Models}

\begin{figure}[t]
\centering
\epsfig{file=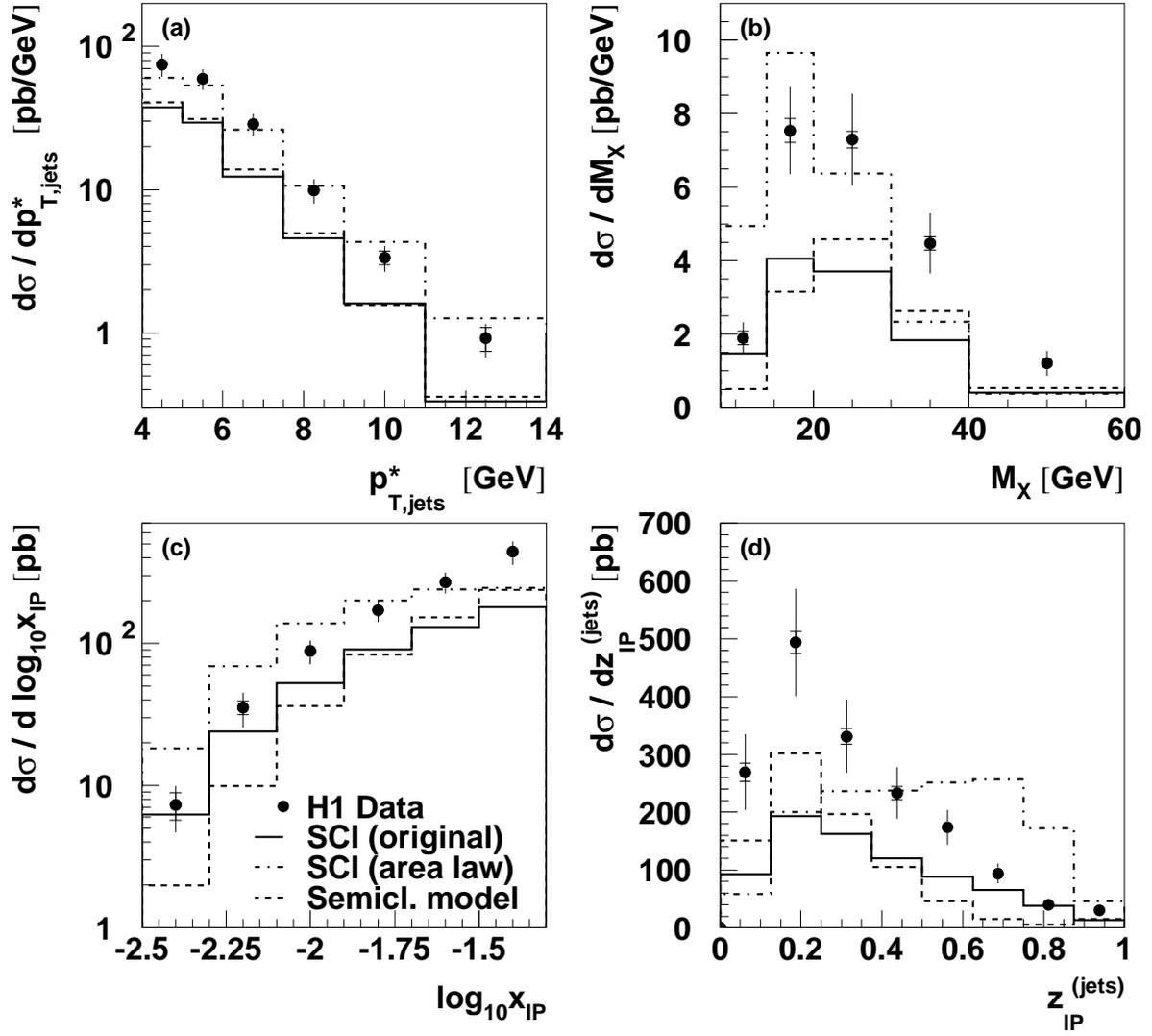,width=1.0\linewidth}
\caption{  Differential dijet cross sections as functions of {\em (a)} 
  $p^*_{T,jets}$, {\em (b)} $M_X$, {\em (c)} $\log x_\pom$ and
  {\em (d)} $z_\pom^{(jets)}$.  The data are compared to the original
  version of the Soft Colour Interaction (SCI) model, labelled `SCI
  (original)', the prediction of the refined SCI version based on a
  generalised area law for string reconnections, labelled `SCI (area
  law)', and to the semiclassical model.}
\label{fig6}
\end{figure}

The Soft Colour Interactions (SCI) and semiclassical models (section
\ref{scisec}) both give a reasonably good description of inclusive
diffraction at HERA with a small number of free parameters.  In
fig.~\ref{fig6}, the predictions of these models are compared with the
dijet cross sections as functions of $p^*_{T,jets}$, $M_X$,
$\log x_\pom$ and $z_\pom^{(jets)}$.  With the exception of the
cross section differential in $M_X$, the data shown are identical to
those in earlier figures.  The original version of SCI gives a
reasonable description of the shapes of the differential distributions
of the dijet data, but the overall cross section is too low by a
factor of about 2.  The refined version of the SCI model, based on a
generalised area law for string rearrangements, gives an improved
description of $F_2^{D(3)}$ at low $Q^2$. It also reproduces the
normalisation of the dijet cross sections much better than the
original version.  However, the shapes of the differential
distributions are not
described, with the exception of $p^*_{T,jets}$.

The semiclassical model gives a good description of the shapes of the
distributions, but the total predicted dijet cross section is only
around half that measured.  The free parameters of the semiclassical
model were determined using only $F_2^{D(3)}$ data in the region
$x_\pom<0.01$. Even at low $x_\pom$, the predictions lie significantly
below the dijet data (fig.~\ref{fig6}c).  It is possible that the
inclusion of NLO terms would improve the description of the data by
the semiclassical model.

\subsection{Colour Dipole and 2-Gluon Exchange Models}
\label{2gluon}

In this section, the saturation and BJLW models (section \ref{ggsec}),
based on the ideas of dipole cross sections and 2-gluon exchange, are
compared with the dijet data. Because of the nature of the 2-gluon
models, only final state parton showers are included in the
simulations.  A restricted data sample with the additional cut
\begin{equation}
x_\pom<0.01
\end{equation}
is studied, because the calculations were carried out under the
assumption of low $x_\pom$ to avoid contributions from secondary
reggeon exchanges and ensure that the proton parton distributions are
gluon dominated.  Applying this additional restriction reduces the
number of events in the data sample by a factor of approximately 4.

The resolved pomeron model implies the presence of a soft pomeron
remnant. The same is true for $q\overline{q}g$ production within the
saturation model where the gluon behaves in a `remnant-like' manner,
due to the $k_T$-ordering condition imposed in the calculations. By
contrast, the $q\overline{q}g$ calculation within the BJLW model
imposes high transverse momenta on all three partons and is not
restricted to $k_T$-ordered configurations.  Any `remnant' system
beyond the dijets in this model is thus expected to have relatively
large $p_T$.  To gain more insight into the properties of the part of
the hadronic final state not belonging to the jets, a new observable
$p_{T,rem}^{(\pom)}$ is introduced. By analogy with the definition
of $E_{rem}^{(\gamma)}$ (section \ref{section:xgam}), this variable
measures the transverse momentum of all hadronic final state particles
in the pomeron hemisphere of the $\gamma^*\pom$ centre-of-mass frame
($\eta^\dagger>0$) not belonging to the two highest $p^*_T$ jets.

\begin{figure}[tb]
\centering
\epsfig{file=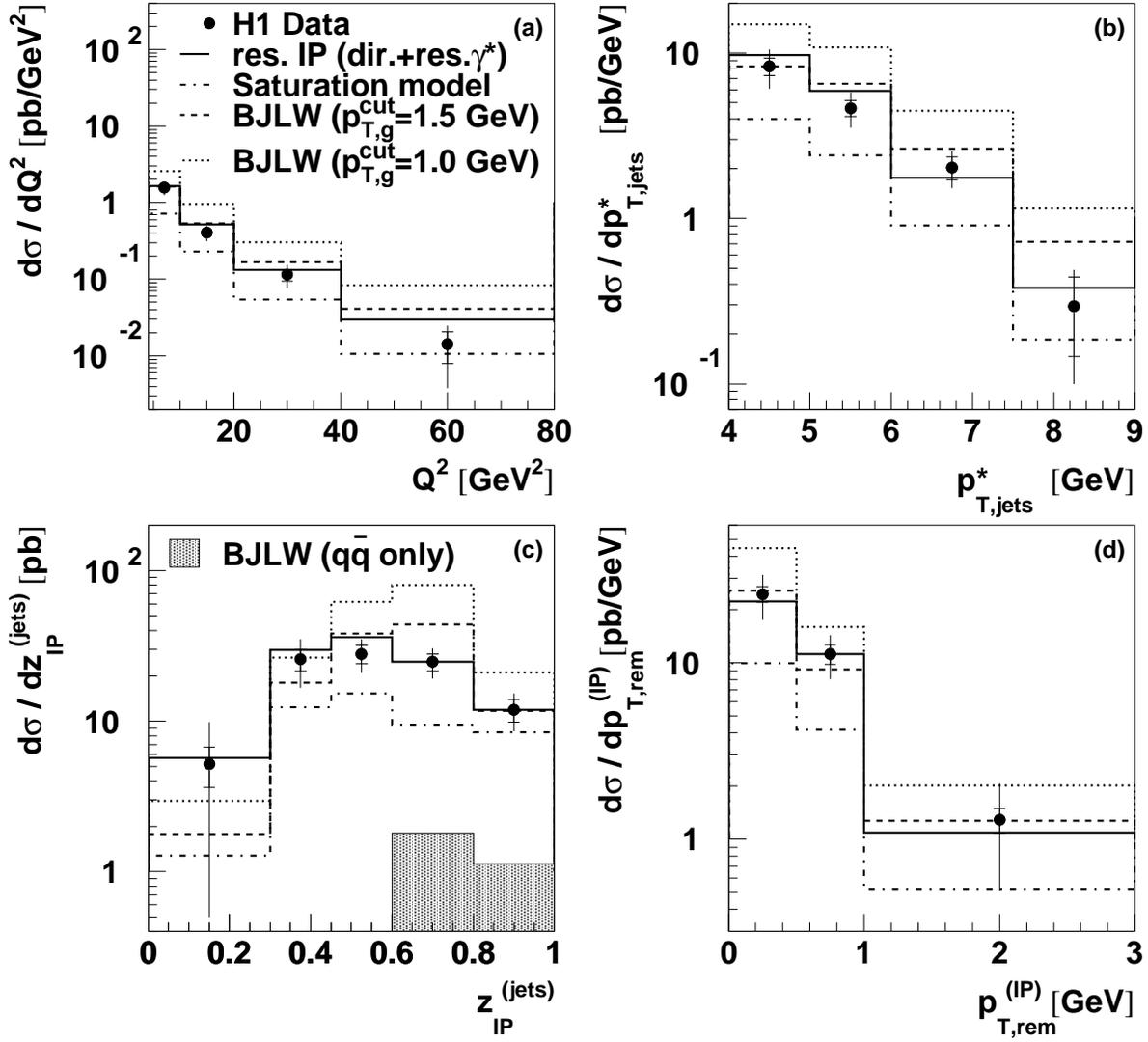,width=1.0\linewidth}
\caption{  Diffractive dijet cross sections in the restricted kinematic range
  $x_\pom<0.01$, shown as functions of {\em (a)} $Q^2$, {\em (b)}
  $p^*_{T,jets}$, {\em (c)} $z_\pom^{(jets)}$ and {\em (d)}
  $p_{T,rem}^{(\pom)}$, the latter denoting the summed transverse
  momentum of the final state particles not belonging to the two
  highest $p^*_T$ jets and located in the pomeron hemisphere of the
  $\gamma^*\pom$ centre-of-mass frame.  The data are compared to the
  saturation, BJLW and resolved pomeron (`fit 2', direct and resolved
  virtual photons) models. For the BJLW model, the contribution from
  $q\overline{q}$ states alone and the sum of the $q\overline{q}$ and
  $q\overline{q}g$ contributions for two different values of the $p_T$
  cut-off for the gluon $p_{T,g}^{cut}$ are shown.}
\label{fig7}
\end{figure}

Dijet cross sections for the region $x_\pom<0.01$ differential in
$Q^2$, $p_{T,jets}^*$, $z^{(jets)}_\pom$ and $p_{T,rem}^{(\pom)}$ are
shown in fig.~\ref{fig7}. They are compared with the predictions of
the saturation, BJLW and resolved pomeron (`fit 2') models.  The
saturation model is able to reproduce the shapes of the measured cross
sections, though the overall predicted dijet rate is too low by a
factor of approximately 2.  The normalisation of the saturation model
is fixed from the fit to inclusive $F_2$ data and by the assumed
$e^{6t}$ dependence for diffractive processes. The total predicted
dijet cross section would increase whilst preserving a good
description of $F_2^{D(3)}$ if the $t$ dependence were found to be
harder for dijet production than for inclusive diffraction.

In the BJLW model, the contribution from $q\overline{q}$ states alone
is negligibly small even at large values of $z_\pom$.  This is in
accordance with the expectation for high $p_T$, high $M_X$ diffractive
final states. The predicted $q\overline{q}g$ contribution is much
larger.  The normalisation of the BJLW model for $q\overline{q}g$
production can be controlled by tuning the lower cut-off on the
transverse momentum of the gluon $p_{T,g}^{cut}$ in the calculations.
If this cut-off is set to $1.5 \rm\ GeV$, the total cross section for
dijet production with $x_\pom<0.01$ is approximately correct in the
model.  Lowering $p_{T,g}^{cut}$ to $1.0 \rm\ GeV$ leads to a
prediction significantly above the measured cross section. The
description of the shapes of the differential cross sections
is reasonable apart
from small discrepancies in the $z_\pom^{(jets)}$ distribution.  The
differences between the predictions of the saturation and BJLW models
may originate from the different parameterisations of
$\mathcal{F}(x,k_T^2)$, the different treatments of non-$k_T$-ordered
configurations or from the assumed $t$ dependences.

The resolved pomeron model, in which the non-$k_T$-ordered resolved
photon contributions are small in the low $\xpom$ region, continues to
give the best description of all observables, including the
$p_{T,rem}^{(\pom)}$ distribution.  The good description of the
$p_{T,rem}^{(\pom)}$ distribution by both the resolved pomeron and the
BJLW models indicates that the present data are not easily able to
discriminate between models with a soft `remnant' and those with a
third high-$p_T$ parton.

\subsection{3-Jet Production}

\begin{figure}[htb]
\centering
\epsfig{file=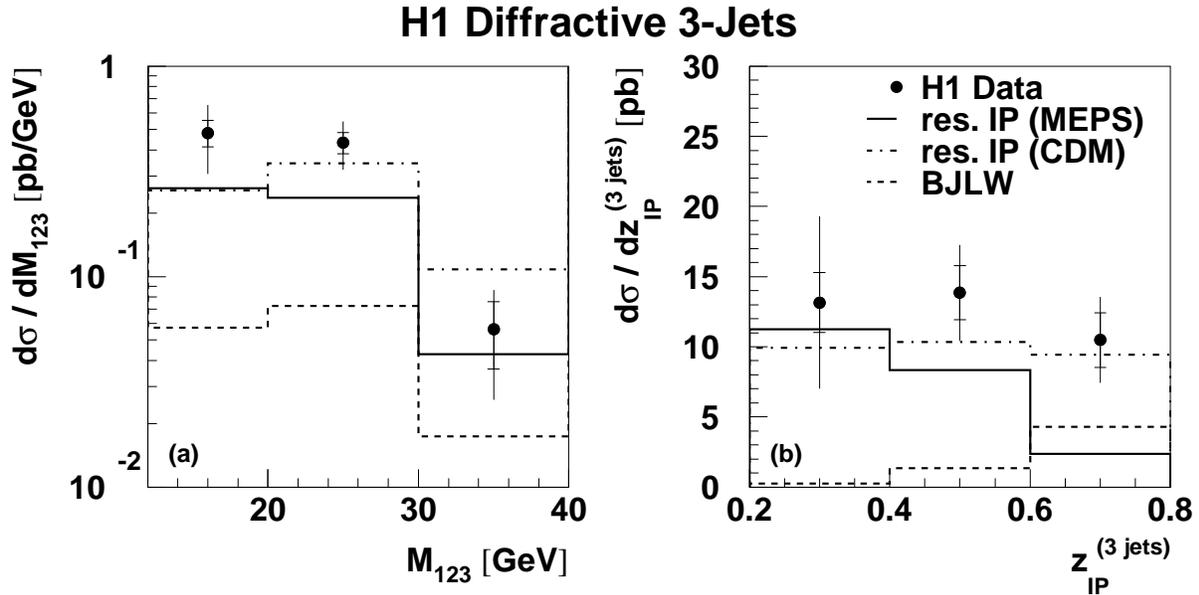,width=1.0\linewidth}
\caption{  Differential cross sections for diffractive 3-jet production
  as functions of {\em (a)} the 3-jet invariant mass $M_{123}$ and
  {\em (b)} the corresponding $z_\pom$-variable $z_\pom^{(3 \ jets)}$,
  measuring the colourless exchange momentum fraction which enters the
  hard interaction. The data are compared with the resolved pomeron
  model with two different approaches for higher order QCD diagrams,
  the parton shower model (labelled `MEPS') and the colour dipole
  approach (labelled `CDM').  The `H1 fit 2' parameterisation is used
  and direct and resolved virtual photon contributions are included.
  The BJLW model is also shown, including $q\overline{q}$ and
  $q\overline{q}g$ contributions, with the cut-off for the gluon
  $p_{T,g}^{cut}$ set to $1.5 \rm\ GeV$.}
\label{fig8}
\end{figure}

The diffractive production of three high-$p_T$ jets as components of
the $X$ system has been investigated.  Except for the requirement on
the number of jets, the analysis is identical to the dijet analysis,
such that no requirements are made on possible hadronic activity
beyond the jets.  In fig.~\ref{fig8}, the measured 3-jet cross
sections are presented as functions of the 3-jet invariant mass
$M_{123}$ and
\begin{equation}
z_\pom^{(3 \ jets)}=\frac{Q^2+M_{123}^2}{Q^2+M_X^2} \ ,
\end{equation}
which, similarly to $z_\pom^{(jets)}$ for dijet events, is a measure
of the fraction of the energy of the $X$ system which is contained in
the jets. The $z_\pom^{(3 \ jets)}$ cross section is measured up to
0.8.  With the present statistics, it is not possible to extract a
cross section for the interesting region $0.8<z_\pom^{(3 \ 
  jets)}\leq1.0$, which corresponds approximately to `exclusive' 3-jet
production.  The measured $z_\pom^{(3 \ jets)}$ cross section
demonstrates that additional hadronic activity beyond the jets is
typically present even in the 3-jet sample.

The data are compared with the resolved pomeron model (`fit 2'), with
the hard interaction evaluated at a scale $\mu^2=Q^2+p_T^2$.  Direct
and resolved $\gamma^*$ contributions are included.  Because the
leading order for 3-parton final states is $\mathcal{O}(\alpha_s^2)$,
two different approximations for higher order QCD diagrams are
considered here, the parton shower model (MEPS) and the colour dipole
approach (CDM).  The measured cross sections are well described when
using CDM. The MEPS simulation tends to lie below the data at low
$M_{123}$ or high $z_\pom^{(3 \ jets)}$.

The BJLW calculation with $p_{T,g}^{cut}= 1.5 \rm\ GeV$ is not able to
accommodate the observed rate of 3-jet events. The predicted cross
section increases towards the high $z_\pom^{(3 \ jets)}$ regime of
exclusive 3-jet production.  For kinematic reasons, the 3-jet sample
originates from the region $x_\pom>0.01$, where contributions from the
proton quark distributions and secondary exchanges, which are not
included in the 2-gluon models, can no longer be neglected. An
improvement in the predictions of dipole models may also come through
the inclusion of higher multiplicity photon fluctuations such as
$q\overline{q}gg$, which have not yet been calculated.


\section{Summary and Final Remarks}
\label{chapter5}

An analysis of the production of jets as components of the
dissociating photon system $X$ in the diffractive DIS reaction
$ep\rightarrow eXY$ has been presented for $4<Q^2<80 \ 
\mathrm{GeV^2}$, $x_\pom<0.05$, $p^*_{T, jet}>4 \ \mathrm{GeV}$,
$M_Y<1.6 \ \mathrm{GeV}$ and $|t|<1.0 \ \mathrm{GeV^2}$. The kinematic
range has been extended to lower $Q^2$ and $p^*_{T,jet}$ compared to previous
measurements \cite{H1:d2j94} and the statistical precision is much
improved. Cross sections for the production of three high transverse
momentum jets have been measured for the first time in
diffraction.

The observed dijet events typically exhibit a structure where, 
in addition to the reconstructed jets, the $X$
system contains hadronic energy with transverse momentum below the jet
scale. The dijet invariant mass
is thus generally smaller than $M_X$.  Viewed in the proton rest
frame, the data clearly require the dominance of higher multiplicity
photon fluctuations (e.g. $q\overline{q}g$) over the simplest
$q\overline{q}$ configuration.  Considered in the proton infinite 
momentum frame,
the data show that the diffractive gluon distribution is much larger
than the quark distribution.

The data can be described by a `resolved partonic pomeron' model, 
with diffractive parton distributions extracted from $F_2^{D(3)}$ data.
The good description from this model
strongly supports the validity of diffractive hard scattering
factorisation in DIS. The
dominant contribution in the model arises from a diffractive exchange
with factorising $x_\pom$ dependence (`Regge' factorisation).  
A value of $\alpha_\pom(0) =
1.17 \ \pm 0.03 \ ({\rm stat.}) \ \pm 0.06 \ ({\rm syst.}) \ 
^{+0.03}_{-0.07} \ ({\rm model}) $ is obtained for the intercept of
the leading trajectory from fits to the dijet data. 
The compatibility of the data with QCD hard scattering and
Regge factorisation contrasts with the observed strong
factorisation breaking when diffractive $ep$ and
$p \bar{p}$ data are compared \cite{facbreak,actw}.
The dijet data give the best constraints to date on the
pomeron gluon distribution.  The data require a large fraction
($80-90\%$, as obtained in \cite{H1:F2d94}) of the pomeron momentum to
be carried by gluons with a momentum distribution which is
comparatively flat in $z_{\pom}$.  
Predictions derived from the `flat
gluon' (or `fit 2') parameterisation in \cite{H1:F2d94}, with higher
order QCD effects modelled using parton showers, are in remarkably
good agreement with all aspects of the dijet data with the single exception
of the $\av{\eta}_{jets}^{lab}$ dependence.
The level of agreement between the resolved pomeron model and the data
is better than that obtained from leading order predictions for
inclusive $ep$ dijet data (e.g. \cite{incljets}), where the NLO
corrections are approximately $40\%$ in a similar region of $Q^2$ and
$p^*_{T,jets}$. 

The two versions of the Soft Colour Interactions (SCI) model are not
able to reproduce the overall dijet rate and the shapes of the
differential cross sections at the same time.  The similarly motivated
semiclassical model in its present (leading order) form achieves a
good description of the shapes of the differential distributions but
underestimates the total dijet cross section.

Models based on colour dipole cross sections and 2-gluon exchange have
been compared with the dijet data in the restricted region
$x_\pom<0.01$.  The saturation model, which takes only $k_T$ ordered
configurations into account, describes the shapes of the jet
distributions but underestimates the overall cross section. The
normalisation of the BJLW model, in which strong $k_T$ ordering is not
imposed, is close to the data if a cut-off for the gluon transverse
momentum of $p^{cut}_{T,g}= 1.5 \rm\ GeV$ is chosen for the
$q \bar{q} g$ contribution.  The shapes of
the differential distributions are reasonably well described.

Strong conclusions cannot yet
be drawn from the 3-jet production cross sections, 
because of the limited statistical accuracy and the
kinematic restriction to large $\xpom$ implied by the requirement of
three high $p_T$ jets. At the present level of precision, the partonic
pomeron predictions based on the `fit 2' parameterisation in
\cite{H1:F2d94} are in good agreement with the 3-jet cross sections,
provided the CDM model of higher order QCD effects is used.  The BJLW
model is unable to reproduce the rate of observed 3-jet events when
$p^{cut}_{T,g}$ is kept fixed at $1.5 \rm\ GeV$.

In conclusion, diffractive jet production has been shown to be a
powerful tool to gain insight into the underlying QCD dynamics of
diffraction, in particular the role of gluons.  The jet cross sections
are sensitive to differences between phenomenological models which all
give a reasonable description of $F_2^{D(3)}$. Models based on
fully factorisable diffractive parton distributions continue to be
successful. Progress in calculations based 
on 2-gluon exchange has led to improved agreement with the data.


\section*{Acknowledgements}

We are grateful to the HERA machine group whose outstanding efforts
have made and continue to make this experiment possible. We thank the
engineers and technicians for their work in constructing and now
maintaining the H1 detector, our funding agencies for financial
support, the DESY technical staff for continual assistance, and the
DESY directorate for the hospitality which they extend to the non DESY
members of the collaboration.
We have benefited from interesting discussions with 
J.~Bartels, A.~Hebecker and M.~McDermott.



\def\capa{ Differential hadron level dijet cross sections.  Here and
elsewhere, the quoted differential cross sections are average values
over the specified intervals.  }

\def\capb{ 
Differential hadron level dijet cross sections (continued).
}

\def\capc{ 
Differential hadron level dijet cross sections in four bins of 
$\log_{10}x_\pom$.
}

\def\capd{ 
Differential hadron level dijet cross sections in four bins of $Q^2+p_T^2$.
}

\def\capf{ 
Differential hadron level dijet cross sections in the restricted kinematical 
range $x_\pom<0.01$.
}

\def\capg{ 
Differential hadron level 3-Jet cross sections.
}

  
 \begin{table}[htb]
 \centering \sf \scriptsize 
 \begin{tabular}
 {|c|d{-1}cd{-1}|d{-1}|d{-1}d{-1}d{-1}|}
 \hline
 \multicolumn{8}{|l|}
{ Dijet cross section as a function of $Q^2                             $.} \\
 \hline
  Bin & \multicolumn{3}{c|}
{$Q^2                              \rm\ [GeV^2                           ]$} & 
 \multicolumn{1}{c|}
{$\sigma \rm\ [pb/GeV^2                           ]$} & 
 \multicolumn{1}{c}
 {stat. err. $[\%]$} & 
 \multicolumn{1}{c}
 {syst. err. $[\%]$} & 
 \multicolumn{1}{c|}
 {tot. err. $[\%]$} \\ 
 \hline
$ 1$&    4.0&--&    6.0&   21.4&  4.5& 16.4& 17.0\\
$ 2$&    6.0&--&   10.0&   13.0&  4.0& 16.1& 16.6\\
$ 3$&   10.0&--&   15.0&    6.3&  4.8& 17.0& 17.7\\
$ 4$&   15.0&--&   20.0&    4.1&  6.2& 16.7& 17.8\\
$ 5$&   20.0&--&   30.0&    2.3&  5.8& 16.4& 17.4\\
$ 6$&   30.0&--&   40.0&    1.2&  8.0& 16.3& 18.2\\
$ 7$&   40.0&--&   50.0&    0.7& 10.4& 19.7& 22.3\\
$ 8$&   50.0&--&   60.0&    0.7& 12.5& 23.9& 27.0\\
$ 9$&   60.0&--&   80.0&    0.4& 11.9& 29.6& 31.9\\
 \hline
 \hline
 \multicolumn{8}{|l|}
{ Dijet cross section as a function of $p^*_{T,jets}                    $.} \\
 \hline
  Bin & \multicolumn{3}{c|}
{$p^*_{T,jets}                     \rm\ [GeV                             ]$} & 
 \multicolumn{1}{c|}
{$\sigma \rm\ [pb/GeV                             ]$} & 
 \multicolumn{1}{c}
 {stat. err. $[\%]$} & 
 \multicolumn{1}{c}
 {syst. err. $[\%]$} & 
 \multicolumn{1}{c|}
 {tot. err. $[\%]$} \\ 
 \hline
$ 1$&    4.0&--&    5.0&   74.9&  4.0& 17.8& 18.3\\
$ 2$&    5.0&--&    6.0&   59.5&  3.3& 16.4& 16.8\\
$ 3$&    6.0&--&    7.5&   28.8&  3.9& 17.3& 17.8\\
$ 4$&    7.5&--&    9.0&    9.9&  7.0& 17.9& 19.2\\
$ 5$&    9.0&--&   11.0&    3.4& 11.0& 17.7& 20.8\\
$ 6$&   11.0&--&   14.0&    0.9& 18.9& 18.6& 26.5\\
 \hline
 \hline
 \multicolumn{8}{|l|}
{ Dijet cross section as a function of $\av{\eta}_{jets}^{lab}          $.} \\
 \hline
  Bin & \multicolumn{3}{c|}
{$\av{\eta}_{jets}^{lab}          $} & 
 \multicolumn{1}{c|}
{$\sigma \rm\ [pb]$} & 
 \multicolumn{1}{c}
 {stat. err. $[\%]$} & 
 \multicolumn{1}{c}
 {syst. err. $[\%]$} & 
 \multicolumn{1}{c|}
 {tot. err. $[\%]$} \\ 
 \hline
$ 1$&  -1.00&--&  -0.66&   22.4& 13.5& 34.3& 36.9\\
$ 2$&  -0.66&--&  -0.33&   68.9&  6.3& 17.7& 18.8\\
$ 3$&  -0.33&--&   0.00&  112.8&  4.7& 15.7& 16.4\\
$ 4$&   0.00&--&   0.33&  131.6&  4.2& 15.7& 16.3\\
$ 5$&   0.33&--&   0.66&  127.9&  4.3& 17.5& 18.0\\
$ 6$&   0.66&--&   1.00&   85.3&  5.1& 17.4& 18.2\\
$ 7$&   1.00&--&   1.50&   16.4&  6.8& 25.6& 26.5\\
 \hline
 \hline
 \multicolumn{8}{|l|}
{ Dijet cross section as a function of $M_X                             $.} \\
 \hline
  Bin & \multicolumn{3}{c|}
{$M_X                              \rm\ [GeV                             ]$} & 
 \multicolumn{1}{c|}
{$\sigma \rm\ [pb/GeV                             ]$} & 
 \multicolumn{1}{c}
 {stat. err. $[\%]$} & 
 \multicolumn{1}{c}
 {syst. err. $[\%]$} & 
 \multicolumn{1}{c|}
 {tot. err. $[\%]$} \\ 
 \hline
$ 1$&    8.0&--&   14.0&    1.9& 10.1& 20.0& 22.4\\
$ 2$&   14.0&--&   20.0&    7.5&  4.4& 15.1& 15.7\\
$ 3$&   20.0&--&   30.0&    7.3&  3.2& 16.9& 17.2\\
$ 4$&   30.0&--&   40.0&    4.5&  4.0& 17.8& 18.3\\
$ 5$&   40.0&--&   60.0&    1.2&  6.2& 27.1& 27.8\\
 \hline
 \hline
 \multicolumn{8}{|l|}
{ Dijet cross section as a function of $W                               $.} \\
 \hline
  Bin & \multicolumn{3}{c|}
{$W                                \rm\ [GeV                             ]$} & 
 \multicolumn{1}{c|}
{$\sigma \rm\ [pb/GeV                             ]$} & 
 \multicolumn{1}{c}
 {stat. err. $[\%]$} & 
 \multicolumn{1}{c}
 {syst. err. $[\%]$} & 
 \multicolumn{1}{c|}
 {tot. err. $[\%]$} \\ 
 \hline
$ 1$&   90.0&--&  115.0&    1.1&  6.6& 20.3& 21.3\\
$ 2$&  115.0&--&  140.0&    1.4&  5.1& 18.5& 19.2\\
$ 3$&  140.0&--&  165.0&    1.7&  4.4& 18.0& 18.5\\
$ 4$&  165.0&--&  190.0&    1.3&  4.5& 17.7& 18.3\\
$ 5$&  190.0&--&  215.0&    1.1&  4.7& 17.7& 18.3\\
$ 6$&  215.0&--&  240.0&    0.9&  5.4& 17.0& 17.8\\
$ 7$&  240.0&--&  260.0&    0.5& 10.3& 28.5& 30.3\\
 \hline
 \end{tabular} \rm \normalsize
 \caption{\capa}
 \label{tab:xsa}
 \end{table}

 \begin{table}[htb]
 \centering \sf \scriptsize 
 \begin{tabular}
 {|c|d{-1}cd{-1}|d{-1}|d{-1}d{-1}d{-1}|}
 \hline
 \multicolumn{8}{|l|}
{ Dijet cross section as a function of $\log_{10}x_\pom                 $.} \\
 \hline
  Bin & \multicolumn{3}{c|}
{$\log_{10}x_\pom                 $} & 
 \multicolumn{1}{c|}
{$\sigma \rm\ [pb]$} & 
 \multicolumn{1}{c}
 {stat. err. $[\%]$} & 
 \multicolumn{1}{c}
 {syst. err. $[\%]$} & 
 \multicolumn{1}{c|}
 {tot. err. $[\%]$} \\ 
 \hline
$ 1$&   -2.5&--&   -2.3&    7.3& 21.8& 28.8& 36.1\\
$ 2$&   -2.3&--&   -2.1&   35.4& 10.8& 25.1& 27.4\\
$ 3$&   -2.1&--&   -1.9&   88.2&  6.8& 17.5& 18.8\\
$ 4$&   -1.9&--&   -1.7&  171.2&  4.7& 16.3& 17.0\\
$ 5$&   -1.7&--&   -1.5&  269.3&  3.6& 16.3& 16.7\\
$ 6$&   -1.5&--&   -1.3&  440.7&  3.2& 18.6& 18.8\\
 \hline
 \hline
 \multicolumn{8}{|l|}
{ Dijet cross section as a function of $\log_{10}\beta                  $.} \\
 \hline
  Bin & \multicolumn{3}{c|}
{$\log_{10}\beta                  $} & 
 \multicolumn{1}{c|}
{$\sigma \rm\ [pb]$} & 
 \multicolumn{1}{c}
 {stat. err. $[\%]$} & 
 \multicolumn{1}{c}
 {syst. err. $[\%]$} & 
 \multicolumn{1}{c|}
 {tot. err. $[\%]$} \\ 
 \hline
$ 1$&   -2.8&--&   -2.5&   24.9& 11.3& 26.4& 28.7\\
$ 2$&   -2.5&--&   -2.2&   88.3&  5.6& 18.1& 19.0\\
$ 3$&   -2.2&--&   -1.9&  129.9&  4.3& 16.7& 17.2\\
$ 4$&   -1.9&--&   -1.6&  152.7&  3.9& 17.4& 17.9\\
$ 5$&   -1.6&--&   -1.3&  145.9&  4.3& 16.8& 17.3\\
$ 6$&   -1.3&--&   -1.1&   85.0&  7.0& 17.5& 18.8\\
$ 7$&   -1.1&--&   -0.8&   53.4&  7.8& 17.4& 19.0\\
$ 8$&   -0.8&--&   -0.5&   13.5& 17.7& 29.8& 34.6\\
 \hline
 \hline
 \multicolumn{8}{|l|}
{ Dijet cross section as a function of $z_\pom^{(jets)}                 $.} \\
 \hline
  Bin & \multicolumn{3}{c|}
{$z_\pom^{(jets)}                 $} & 
 \multicolumn{1}{c|}
{$\sigma \rm\ [pb]$} & 
 \multicolumn{1}{c}
 {stat. err. $[\%]$} & 
 \multicolumn{1}{c}
 {syst. err. $[\%]$} & 
 \multicolumn{1}{c|}
 {tot. err. $[\%]$} \\ 
 \hline
$ 1$&  0.000&--&  0.125&  269.4&  5.8& 23.7& 24.4\\
$ 2$&  0.125&--&  0.250&  493.9&  3.8& 18.4& 18.8\\
$ 3$&  0.250&--&  0.375&  331.3&  4.2& 18.6& 19.1\\
$ 4$&  0.375&--&  0.500&  233.2&  4.9& 18.5& 19.2\\
$ 5$&  0.500&--&  0.625&  174.2&  5.9& 16.1& 17.2\\
$ 6$&  0.625&--&  0.750&   94.0&  8.1& 16.3& 18.2\\
$ 7$&  0.750&--&  0.875&   39.8& 11.7& 16.3& 20.0\\
$ 8$&  0.875&--&  1.000&   30.0& 16.7& 24.5& 29.7\\
 \hline
 \hline
 \multicolumn{8}{|l|}
{ Dijet cross section as a function of $x_\gamma^{(jets)}               $.} \\
 \hline
  Bin & \multicolumn{3}{c|}
{$x_\gamma^{(jets)}               $} & 
 \multicolumn{1}{c|}
{$\sigma \rm\ [pb]$} & 
 \multicolumn{1}{c}
 {stat. err. $[\%]$} & 
 \multicolumn{1}{c}
 {syst. err. $[\%]$} & 
 \multicolumn{1}{c|}
 {tot. err. $[\%]$} \\ 
 \hline
$ 1$&    0.0&--&    0.2&   25.4& 14.3& 35.1& 37.9\\
$ 2$&    0.2&--&    0.4&  104.8&  6.5& 17.7& 18.9\\
$ 3$&    0.4&--&    0.6&  153.8&  5.0& 18.1& 18.8\\
$ 4$&    0.6&--&    0.8&  331.5&  3.6& 18.0& 18.3\\
$ 5$&    0.8&--&    1.0&  428.3&  3.1& 16.7& 17.0\\
 \hline
 \hline
 \multicolumn{8}{|l|}
{ Dijet cross section as a function of $E_{rem}^{(\gamma)}              $.} \\
 \hline
  Bin & \multicolumn{3}{c|}
{$E_{rem}^{(\gamma)}               \rm\ [GeV                             ]$} & 
 \multicolumn{1}{c|}
{$\sigma \rm\ [pb/GeV                             ]$} & 
 \multicolumn{1}{c}
 {stat. err. $[\%]$} & 
 \multicolumn{1}{c}
 {syst. err. $[\%]$} & 
 \multicolumn{1}{c|}
 {tot. err. $[\%]$} \\ 
 \hline
$ 1$&    0.0&--&    4.0&   38.1&  2.5& 17.2& 17.3\\
$ 2$&    4.0&--&    8.0&    9.0&  4.7& 17.6& 18.2\\
$ 3$&    8.0&--&   12.0&    4.0&  6.7& 25.4& 26.3\\
$ 4$&   12.0&--&   20.0&    2.0&  8.0& 38.6& 39.4\\
 \hline
 \end{tabular} \rm \normalsize
 \caption{\capb}
 \label{tab:xsb}
 \end{table}

 \begin{table}[htb]
 \centering \sf \scriptsize 
 \begin{tabular}
 {|c|d{-1}cd{-1}|d{-1}|d{-1}d{-1}d{-1}|}
 \hline
 \multicolumn{8}{|l|}
{ Dijet cross section as a function of $z_\pom^{(jets)}                 $ for $-1.5<\log_{10}x_\pom<-1.3$.} \\
 \hline
  Bin & \multicolumn{3}{c|}
{$z_\pom^{(jets)}                 $} & 
 \multicolumn{1}{c|}
{$\sigma \rm\ [pb]$} & 
 \multicolumn{1}{c}
 {stat. err. $[\%]$} & 
 \multicolumn{1}{c}
 {syst. err. $[\%]$} & 
 \multicolumn{1}{c|}
 {tot. err. $[\%]$} \\ 
 \hline
$ 1$&   0.00&--&   0.15&  232.9&  6.0& 29.7& 30.3\\
$ 2$&   0.15&--&   0.30&  209.4&  5.3& 24.2& 24.8\\
$ 3$&   0.30&--&   0.50&   85.4&  6.4& 20.8& 21.8\\
$ 4$&   0.50&--&   0.70&   30.9& 10.4& 18.8& 21.5\\
$ 5$&   0.70&--&   1.00&    3.4& 28.9& 47.0& 55.1\\
 \hline
 \hline
 \multicolumn{8}{|l|}
{ Dijet cross section as a function of $z_\pom^{(jets)}                 $ for $-1.75<\log_{10}x_\pom<-1.5$.} \\
 \hline
  Bin & \multicolumn{3}{c|}
{$z_\pom^{(jets)}                 $} & 
 \multicolumn{1}{c|}
{$\sigma \rm\ [pb]$} & 
 \multicolumn{1}{c}
 {stat. err. $[\%]$} & 
 \multicolumn{1}{c}
 {syst. err. $[\%]$} & 
 \multicolumn{1}{c|}
 {tot. err. $[\%]$} \\ 
 \hline
$ 1$&    0.0&--&    0.2&   97.1&  6.4& 20.8& 21.8\\
$ 2$&    0.2&--&    0.4&  134.3&  5.3& 19.0& 19.7\\
$ 3$&    0.4&--&    0.6&   63.4&  7.1& 16.3& 17.7\\
$ 4$&    0.6&--&    0.8&   21.8& 12.6& 16.6& 20.8\\
$ 5$&    0.8&--&    1.0&    8.5& 25.8& 34.4& 43.0\\
 \hline
 \hline
 \multicolumn{8}{|l|}
{ Dijet cross section as a function of $z_\pom^{(jets)}                 $ for $-2.0<\log_{10}x_\pom<-1.75$.} \\
 \hline
  Bin & \multicolumn{3}{c|}
{$z_\pom^{(jets)}                 $} & 
 \multicolumn{1}{c|}
{$\sigma \rm\ [pb]$} & 
 \multicolumn{1}{c}
 {stat. err. $[\%]$} & 
 \multicolumn{1}{c}
 {syst. err. $[\%]$} & 
 \multicolumn{1}{c|}
 {tot. err. $[\%]$} \\ 
 \hline
$ 1$&   0.00&--&   0.30&   37.8&  8.3& 23.4& 24.9\\
$ 2$&   0.30&--&   0.45&   59.7&  9.2& 18.2& 20.4\\
$ 3$&   0.45&--&   0.60&   49.1& 11.2& 19.4& 22.4\\
$ 4$&   0.60&--&   0.80&   33.8& 10.9& 19.1& 22.0\\
$ 5$&   0.80&--&   1.00&    9.0& 21.8& 24.8& 33.0\\
 \hline
 \hline
 \multicolumn{8}{|l|}
{ Dijet cross section as a function of $z_\pom^{(jets)}                 $ for $\log_{10}x_\pom<-2.0$.} \\
 \hline
  Bin & \multicolumn{3}{c|}
{$z_\pom^{(jets)}                 $} & 
 \multicolumn{1}{c|}
{$\sigma \rm\ [pb]$} & 
 \multicolumn{1}{c}
 {stat. err. $[\%]$} & 
 \multicolumn{1}{c}
 {syst. err. $[\%]$} & 
 \multicolumn{1}{c|}
 {tot. err. $[\%]$} \\ 
 \hline
$ 1$&   0.00&--&   0.30&    5.2& 30.2& 85.2& 90.4\\
$ 2$&   0.30&--&   0.45&   25.8& 16.2& 31.5& 35.4\\
$ 3$&   0.45&--&   0.60&   28.0& 13.9& 21.3& 25.5\\
$ 4$&   0.60&--&   0.80&   24.8& 12.8& 18.9& 22.9\\
$ 5$&   0.80&--&   1.00&   11.9& 17.4& 22.0& 28.0\\
 \hline
 \end{tabular} \rm \normalsize
 \caption{\capc}
 \label{tab:xsc}
 \end{table}

 \begin{table}[htb]
 \centering \sf \scriptsize 
 \begin{tabular}
 {|c|d{-1}cd{-1}|d{-1}|d{-1}d{-1}d{-1}|}
 \hline
 \multicolumn{8}{|l|}
{ Dijet cross section as a function of $z_\pom^{(jets)}                 $ for $20 \ {\rm GeV^2} <Q^2+p_T^2<35 \rm\ GeV^2$.} \\
 \hline
  Bin & \multicolumn{3}{c|}
{$z_\pom^{(jets)}                 $} & 
 \multicolumn{1}{c|}
{$\sigma \rm\ [pb]$} & 
 \multicolumn{1}{c}
 {stat. err. $[\%]$} & 
 \multicolumn{1}{c}
 {syst. err. $[\%]$} & 
 \multicolumn{1}{c|}
 {tot. err. $[\%]$} \\ 
 \hline
$ 1$&    0.0&--&    0.2&  150.5&  6.7& 32.4& 33.1\\
$ 2$&    0.2&--&    0.4&  109.0&  7.3& 26.6& 27.6\\
$ 3$&    0.4&--&    0.6&   45.2& 10.8& 28.8& 30.8\\
$ 4$&    0.6&--&    0.8&   18.7& 16.2& 31.6& 35.5\\
$ 5$&    0.8&--&    1.0&    5.9& 31.6& 54.4& 63.0\\
 \hline
 \hline
 \multicolumn{8}{|l|}
{ Dijet cross section as a function of $z_\pom^{(jets)}                 $ for $35 \ {\rm GeV^2} <Q^2+p_T^2<45 \rm\ GeV^2$.} \\
 \hline
  Bin & \multicolumn{3}{c|}
{$z_\pom^{(jets)}                 $} & 
 \multicolumn{1}{c|}
{$\sigma \rm\ [pb]$} & 
 \multicolumn{1}{c}
 {stat. err. $[\%]$} & 
 \multicolumn{1}{c}
 {syst. err. $[\%]$} & 
 \multicolumn{1}{c|}
 {tot. err. $[\%]$} \\ 
 \hline
$ 1$&    0.0&--&    0.2&   89.7&  7.1& 25.2& 26.2\\
$ 2$&    0.2&--&    0.4&   71.8&  6.9& 21.8& 22.9\\
$ 3$&    0.4&--&    0.6&   39.3&  9.0& 26.1& 27.6\\
$ 4$&    0.6&--&    0.8&   16.9& 14.4& 26.3& 30.0\\
$ 5$&    0.8&--&    1.0&    4.3& 27.7& 26.2& 38.1\\
 \hline
 \hline
 \multicolumn{8}{|l|}
{ Dijet cross section as a function of $z_\pom^{(jets)}                 $ for $45 \ {\rm GeV^2} <Q^2+p_T^2<60 \rm\ GeV^2$.} \\
 \hline
  Bin & \multicolumn{3}{c|}
{$z_\pom^{(jets)}                 $} & 
 \multicolumn{1}{c|}
{$\sigma \rm\ [pb]$} & 
 \multicolumn{1}{c}
 {stat. err. $[\%]$} & 
 \multicolumn{1}{c}
 {syst. err. $[\%]$} & 
 \multicolumn{1}{c|}
 {tot. err. $[\%]$} \\ 
 \hline
$ 1$&    0.0&--&    0.2&   74.6&  7.7& 24.6& 25.8\\
$ 2$&    0.2&--&    0.4&   78.0&  6.7& 24.5& 25.4\\
$ 3$&    0.4&--&    0.6&   43.2&  8.6& 18.6& 20.5\\
$ 4$&    0.6&--&    0.8&   14.7& 14.7& 20.1& 25.0\\
$ 5$&    0.8&--&    1.0&    5.5& 23.6& 28.7& 37.2\\
 \hline
 \hline
 \multicolumn{8}{|l|}
{ Dijet cross section as a function of $z_\pom^{(jets)}                 $ for $Q^2+p_T^2> 60 \ {\rm GeV^2}$.} \\
 \hline
  Bin & \multicolumn{3}{c|}
{$z_\pom^{(jets)}                 $} & 
 \multicolumn{1}{c|}
{$\sigma \rm\ [pb]$} & 
 \multicolumn{1}{c}
 {stat. err. $[\%]$} & 
 \multicolumn{1}{c}
 {syst. err. $[\%]$} & 
 \multicolumn{1}{c|}
 {tot. err. $[\%]$} \\ 
 \hline
$ 1$&    0.0&--&    0.2&   58.7&  9.2& 25.2& 26.9\\
$ 2$&    0.2&--&    0.4&  114.6&  5.8& 17.4& 18.4\\
$ 3$&    0.4&--&    0.6&   73.4&  6.8& 15.6& 17.0\\
$ 4$&    0.6&--&    0.8&   45.3&  9.4& 15.3& 18.0\\
$ 5$&    0.8&--&    1.0&   14.4& 18.3& 22.3& 28.8\\
 \hline
 \end{tabular} \rm \normalsize
 \caption{\capd}
 \label{tab:xsd}
 \end{table}

 \begin{table}[htb]
 \centering \sf \scriptsize 
 \begin{tabular}
 {|c|d{-1}cd{-1}|d{-1}|d{-1}d{-1}d{-1}|}
 \hline
 \multicolumn{8}{|l|}
{ Dijet cross section as a function of $Q^2                             $ for $x_\pom<0.01$.} \\
 \hline
  Bin & \multicolumn{3}{c|}
{$Q^2                              \rm\ [GeV^2                           ]$} & 
 \multicolumn{1}{c|}
{$\sigma \rm\ [pb/GeV^2                           ]$} & 
 \multicolumn{1}{c}
 {stat. err. $[\%]$} & 
 \multicolumn{1}{c}
 {syst. err. $[\%]$} & 
 \multicolumn{1}{c|}
 {tot. err. $[\%]$} \\ 
 \hline
$ 1$&    4.0&--&   10.0&   1.58&  9.7& 18.5& 20.9\\
$ 2$&   10.0&--&   20.0&   0.40& 13.9& 18.4& 23.1\\
$ 3$&   20.0&--&   40.0&   0.12& 17.7& 29.2& 34.1\\
$ 4$&   40.0&--&   80.0&   0.01& 44.7& 58.3& 73.5\\
 \hline
 \hline
 \multicolumn{8}{|l|}
{ Dijet cross section as a function of $p^*_{T,jets}                    $ for $x_\pom<0.01$.} \\
 \hline
  Bin & \multicolumn{3}{c|}
{$p^*_{T,jets}                     \rm\ [GeV                             ]$} & 
 \multicolumn{1}{c|}
{$\sigma \rm\ [pb/GeV                             ]$} & 
 \multicolumn{1}{c}
 {stat. err. $[\%]$} & 
 \multicolumn{1}{c}
 {syst. err. $[\%]$} & 
 \multicolumn{1}{c|}
 {tot. err. $[\%]$} \\ 
 \hline
$ 1$&    4.0&--&    5.0&    8.3& 12.0& 23.9& 26.8\\
$ 2$&    5.0&--&    6.0&    4.7& 11.0& 21.3& 24.0\\
$ 3$&    6.0&--&    7.5&    2.0& 16.0& 19.1& 24.9\\
$ 4$&    7.5&--&    9.0&    0.3& 50.0& 43.2& 66.1\\
 \hline
 \hline
 \multicolumn{8}{|l|}
{ Dijet cross section as a function of $z_\pom^{(jets)}                 $ for $x_\pom<0.01$.} \\
 \hline
  Bin & \multicolumn{3}{c|}
{$z_\pom^{(jets)}                 $} & 
 \multicolumn{1}{c|}
{$\sigma \rm\ [pb]$} & 
 \multicolumn{1}{c}
 {stat. err. $[\%]$} & 
 \multicolumn{1}{c}
 {syst. err. $[\%]$} & 
 \multicolumn{1}{c|}
 {tot. err. $[\%]$} \\ 
 \hline
$ 1$&   0.00&--&   0.30&    5.2& 30.2& 85.2& 90.4\\
$ 2$&   0.30&--&   0.45&   25.8& 16.2& 31.5& 35.4\\
$ 3$&   0.45&--&   0.60&   28.0& 13.9& 21.3& 25.5\\
$ 4$&   0.60&--&   0.80&   24.8& 12.8& 18.9& 22.9\\
$ 5$&   0.80&--&   1.00&   11.9& 17.4& 22.0& 28.0\\
 \hline
 \hline
 \multicolumn{8}{|l|}
{ Dijet cross section as a function of $p_{T,rem}^{(\pom)}              $ for $x_\pom<0.01$.} \\
 \hline
  Bin & \multicolumn{3}{c|}
{$p_{T,rem}^{(\pom)}               \rm\ [GeV                             ]$} & 
 \multicolumn{1}{c|}
{$\sigma \rm\ [pb/GeV                             ]$} & 
 \multicolumn{1}{c}
 {stat. err. $[\%]$} & 
 \multicolumn{1}{c}
 {syst. err. $[\%]$} & 
 \multicolumn{1}{c|}
 {tot. err. $[\%]$} \\ 
 \hline
$ 1$&    0.0&--&    0.5&   24.5& 10.3& 26.8& 28.8\\
$ 2$&    0.5&--&    1.0&   11.2& 12.7& 24.8& 27.9\\
$ 3$&    1.0&--&    3.0&    1.3& 16.0& 58.1& 60.3\\
 \hline
 \end{tabular} \rm \normalsize
 \caption{\capf}
 \label{tab:xsf}
 \end{table}

 \begin{table}[htb]
 \centering \sf \scriptsize 
 \begin{tabular}
 {|c|d{-1}cd{-1}|d{-1}|d{-1}d{-1}d{-1}|}
 \hline
 \multicolumn{8}{|l|}
{ 3-jet cross section as a function of $M_{123}                         $.} \\
 \hline
  Bin & \multicolumn{3}{c|}
{$M_{123}                          \rm\ [GeV                             ]$} & 
 \multicolumn{1}{c|}
{$\sigma \rm\ [pb/GeV                             ]$} & 
 \multicolumn{1}{c}
 {stat. err. $[\%]$} & 
 \multicolumn{1}{c}
 {syst. err. $[\%]$} & 
 \multicolumn{1}{c|}
 {tot. err. $[\%]$} \\ 
 \hline
$ 1$&   12.0&--&   20.0&   0.48& 14.4& 33.0& 36.1\\
$ 2$&   20.0&--&   30.0&   0.43& 11.6& 23.0& 25.7\\
$ 3$&   30.0&--&   40.0&   0.06& 35.3& 40.5& 53.7\\
 \hline
 \hline
 \multicolumn{8}{|l|}
{ 3-jet cross section as a function of $z_\pom^{(3 \ jets)}             $.} \\
 \hline
  Bin & \multicolumn{3}{c|}
{$z_\pom^{(3 \ jets)}             $} & 
 \multicolumn{1}{c|}
{$\sigma \rm\ [pb]$} & 
 \multicolumn{1}{c}
 {stat. err. $[\%]$} & 
 \multicolumn{1}{c}
 {syst. err. $[\%]$} & 
 \multicolumn{1}{c|}
 {tot. err. $[\%]$} \\ 
 \hline
$ 1$&    0.2&--&    0.4&   13.1& 16.2& 43.8& 46.7\\
$ 2$&    0.4&--&    0.6&   13.9& 13.9& 20.5& 24.7\\
$ 3$&    0.6&--&    0.8&   10.5& 18.6& 22.6& 29.2\\
 \hline
 \end{tabular} \rm \normalsize
 \caption{\capg}
 \label{tab:xsg}
 \end{table}



\begin{thebibliography}{99}

\bibitem{obsdiff} ZEUS Collaboration, M.~Derrick {\em et al.},
\Journal{\PLB}{315}{1993}{481}. \\
H1 Collaboration, T.~Ahmed {\em et al.}, \Journal{\NPB}{429}{1994}{477}.

\bibitem{H1:F2d93} H1 Collaboration, T.~Ahmed {\em et al.}, 
\Journal{\PLB}{348}{1995}{681}.

\bibitem{H1:F2d94} H1 Collaboration, C.~Adloff {\em et al.}, 
\Journal{\ZPC}{76}{1997}{613}.

\bibitem{f2dzeus} 
ZEUS Collaboration, J.~Breitweg {\em et al.}, \Journal{\EJC}{1}{1998}{81}; \\
ZEUS Collaboration, J.~Breitweg {\em et al.}, \Journal{\EJC}{6}{1999}{43}.

\bibitem{H1:diffhfs}
H1 Collaboration, C.~Adloff {\em et al.}, \Journal{\EJC}{1}{1998}{495};  \\
H1 Collaboration, C.~Adloff {\em et al.}, \Journal{\PLB}{428}{1998}{206};  \\
H1 Collaboration, C.~Adloff {\em et al.}, \Journal{\EJC}{5}{1998}{439}.  

\bibitem{diffhfs-zeus} ZEUS Collaboration, J.~Breitweg {\em et al.},
\Journal{\PLB}{421}{1998}{368}. 

\bibitem{djzeus} ZEUS Collaboration, J.~Breitweg {\em et al.}, 
\Journal{\EJC}{5}{1998}{41}. 

\bibitem{H1:d2j94} H1 Collaboration, C.~Adloff {\em et al.}, 
\Journal{\EJC}{6}{1999}{421}. 

\bibitem{gg1} M.~Ryskin, {\em Sov. J. Nucl. Phys.} {\bf 52} (1990) 529; \\
              N.~Nikolaev, B.~Zakharov, \Journal{\ZPC}{53}{1992}{331}.

\bibitem{gg} A.~Mueller, \Journal{\NPB}{335}{1990}{115}; \\
             M.~Diehl, \Journal{\ZPC}{66}{1995}{181}.
             
\bibitem{bartelsqq} J.~Bartels, H.~Lotter,  M.~W\"{u}sthoff, \
\Journal{\PLB}{379}{1996}{239} and erratum-ibid. {\bf B 382} (1996) 449; \\
J.~Bartels, C.~Ewerz, H.~Lotter, M.~W\"{u}sthoff, 
\Journal{\PLB}{386}{1996}{389}; \\
H.~Lotter, \Journal{\PLB}{406}{1997}{171}. 

\bibitem{bartelsqqg} J.~Bartels, H.~Jung, M.~W\"{u}sthoff, 
\Journal{\EJC}{11}{1999}{111}; \\
J.~Bartels, H.~Jung, A.~Kyrieleis: {\em Massive $c \bar{c}g$-Production in 
Diffractive DIS}, hep-ph/0010300.

\bibitem{difjetsUA8}
UA8 Collaboration, R.~Bonino {\em et al.}, \Journal{\PLB}{211}{1988}{239}; \\
UA8 Collaboration, A.~Brandt {\em et al.}, \Journal{\PLB}{297}{1992}{417}. 

\bibitem{difjetsCDF} 
CDF Collaboration, F.~Abe {\em et al.}, \Journal{\PRL}{79}{1998}{2636}; \\ 
CDF Collaboration, F.~Abe {\em et al.}, \Journal{\PRL}{80}{1998}{1156}.

\bibitem{facbreak}
CDF Collaboration, T.~Affolder {\em et al.}, \Journal{\PRL}{84}{2000}{5043}.

\bibitem{difjetsD0} 
D0 Collaboration, B.~Abbott {\em et al.}: {\em Hard Single Diffraction in 
$p\bar{p}$ Collisions at 630 and 1800 GeV}, hep-ex/9912061,
submitted to {\PRL}

\bibitem{collins} J.~Collins, \Journal{\PRD}{57}{1998}{3051}
and erratum-ibid. {\bf D 61} (2000) 019902. 

\bibitem{IngSchl} G.~Ingelman, P.~Schlein, \Journal{\PLB}{152}{1985}{256}.

\bibitem{sat}
K.~Golec-Biernat, M.~W\"{u}sthoff, \Journal{\PRD}{59}{1999}{014017}; \\
K.~Golec-Biernat, M.~W\"{u}sthoff, \Journal{\PRD}{60}{1999}{114023}. 

\bibitem{lownussinov}  F.~Low, \Journal{\PRD}{12}{1975}{163}; \\
S.~Nussinov, \Journal{\PRL}{34}{1975}{1286}. 

\bibitem{semicl} W.~Buchm\"{u}ller, T.~Gehrmann, A.~Hebecker, 
\Journal{\NPB}{537}{1999}{477}.

\bibitem{sci} A.~Edin, G.~Ingelman, J.~Rathsman, 
\Journal{\PLB}{366}{1996}{371}; \\
A.~Edin, G.~Ingelman, J.~Rathsman, \Journal{\ZPC}{75}{1997}{57}. 

\bibitem{scinew} J.~Rathsman, \Journal{\PLB}{452}{1999}{364}.

\bibitem{AJM} J.~Bjorken, J.~Kogut, \Journal{\PRD}{8}{1973}{1341}. \\
G.~Bertsch, S.~Brodsky, A.~Goldhaber, J.~Gunion, 
\Journal{\PRL}{47}{1981}{297}.

\bibitem{bmh}
W.~Buchm\"uller, A.~Hebecker, M.~McDermott,
\Journal{\PLB}{410}{1997}{304}.

\bibitem{trentadue} L.~Trentadue, G.~Veneziano, \Journal{\PLB}{323}{1994}{201}.
 
\bibitem{berera} A.~Berera, D.~Soper, \Journal{\PRD}{50}{1994}{4328}.

\bibitem{facproof} M.~Grazzini, L.~Trentadue, G.~Veneziano, 
\Journal{\NPB}{519}{1998}{394}.

\bibitem{dglap} V.~Gribov, L.~Lipatov, 
\Journal{{\em Sov. J. Nucl. Phys.}}{15}{1972}{438, 675}; \\
Y.~Dokshitzer, \Journal{\em Sov. Phys. JETP}{46}{1977}{641}; \\
G.~Altarelli, G.~Parisi, \Journal{\NPB}{126}{1977}{298}.

\bibitem{hautmann} F.~Hautmann, Z.~Kunszt, D.~Soper, 
\Journal{\PRL}{81}{1998}{3333}.

\bibitem{DL:stot} A.~Donnachie, P.~Landshoff, 
\Journal{\PLB}{296}{1992}{227}; \\
J.~Cudell, K.~Kang, S.~Kim, \Journal{\PLB}{395}{1997}{311}.

\bibitem{nikzak90} N.~Nikolaev, B.~Zakharov, \Journal{\ZPC}{49}{1990}{607}. 

\bibitem{mcdermott00} M.~McDermott, {\em The dipole picture of
small-$x$ physics}, hep-ph/0008260. 

\bibitem{bekw} J.~Bartels, J.~Ellis, H.~Kowalski, M.~W\"{u}sthoff, 
\Journal{\EJC}{7}{1999}{443}.

\bibitem{bfkl} 
E.~Kuraev, L.~Lipatov, V.~Fadin, {\em Sov. Phys. JETP} {\bf 45} (1977) 199; \\
Y.~Balitskii, L.~Lipatov, {\em Sov. J. Nucl. Phys.} {\bf 28} (1978) 822; \\
L.~Lipatov, {\em Sov. Phys. JETP} {\bf 63} (1986) 904. 

\bibitem{mcdermott} 
J.~Forshaw, G.~Kerley, G.~Shaw, \Journal{\PRD}{60}{1999}{074012}; \\
L.~Frankfurt, V.~Guzey, M.~McDermott, M.~Strikman:
{\em Unitarity and the QCD-improved dipole picture}, hep-ph/9912547. 

\bibitem{GRVfgluon} M.~Gl\"uck, E.~Reya, A.~Vogt, 
\Journal{\ZPC}{67}{1995}{433}.

\bibitem{dl:form} A.~Donnachie, P.~Landshoff, \Journal{\NPB}{244}{1984}{322}.

\bibitem{hebecker} A.~Hebecker, \Journal{\NPB}{505}{1997}{349}.

\bibitem{rapgap} H.~Jung,  \Journal{\CPC}{86}{1995}{147}. \\
(see also http://www.desy.de/$\sim$jung/rapgap.html)

\bibitem{owens} J.~Owens, \Journal{\PRD}{30}{1984}{943}.

\bibitem{nikolaev} N.\ Nikolaev: {\em Intrinsic $k_{\perp}$ in the Pomeron} 
in {\em Monte Carlo Generators for HERA Physics}, A.~Doyle, G.~Grindhammer, 
G.~Ingelman, H.~Jung (eds.), DESY-PROC-1999-02 (1999) 377.

\bibitem{meps} M.~Bengtsson, T.~Sj\"ostrand, \Journal{\ZPC}{37}{1988}{465}. 

\bibitem{cdm}
G.~Gustafson,  \Journal{\PLB}{175}{1986}{453}; \\
G.~Gustafson, U.~Petterson, \Journal{\NPB}{306}{1988}{746}; \\
B.~Andersson, G.~Gustafson, L.~L\"onnblad, U.~Petterson, 
    \Journal{\ZPC}{43}{1989}{625}; \\
B.~Andersson, G.~Gustafson, L.~L\"onnblad,
    \Journal{\NPB}{339}{1990}{393}.

\bibitem{ariadne} L.~L\"onnblad, \Journal{\CPC}{71}{1992}{15}.

\bibitem{lund} T.~Sj\"ostrand, \Journal{\CPC}{82}{1994}{74}.

\bibitem{heracles} A.~Kwiatkowski, H.~Spiesberger, H.~M\"{o}hring, 
\Journal{\CPC}{69}{1992}{155}. 

\bibitem{sas} 
G.~Schuler, T.~Sj\"ostrand, \Journal{\ZPC}{68}{1995}{607}; \\
G.~Schuler, T.~Sj\"ostrand, \Journal{\PLB}{376}{1996}{193}.

\bibitem{h1:virtgam} H1 Collaboration, C.~Adloff {\em et al.}, 
\Journal{\EJC}{13}{2000}{397}. 

\bibitem{lepto} A.~Edin, G.~Ingelman, J.~Rathsman, 
\Journal{\CPC}{101}{1997}{108}. 

\bibitem{fpschill} F.-P.~Schilling: {\em Diffractive Jet Production in
Deep-Inelastic $e^+ p$ Collisions at HERA}, Ph.D. Thesis, 
University of Heidelberg (Germany), in preparation. Will be available
from http://www-h1.desy.de/publications/theses\_list.html.

\bibitem{H1:det} H1 Collaboration, I.~Abt {\em et al.}, 
\Journal{\NIMA}{386}{1997}{310 and 348}.  

\bibitem{fscomb} H1 Collaboration, C.~Adloff {\em et al.}, 
\Journal{\ZPC}{74}{1997}{221}. 

\bibitem{cdfcone} CDF Collaboration, F.~Abe {\em et al.},  
\Journal{\PRD}{45}{1992}{1448}.  

\bibitem{diffvm} B.~List: {\em Diffraktive $J/\psi$-Produktion in 
Elektron-Proton-St\"o{\ss}en am Speicherring HERA}, Diploma Thesis, 
Tech. Univ. Berlin (1993), unpublished; \\
B.~List, A.~Mastroberardino: {\em DIFFVM: A Monte Carlo Generator for
diffractive processes in ep scattering} in 
{\em Monte Carlo Generators for HERA Physics}, A.~Doyle, G.~Grindhammer, 
G.~Ingelman, H.~Jung (eds.), DESY-PROC-1999-02 (1999) 396.

\bibitem{hztool} J.~Bromley {\em et al.}: {\em HzTool: A Package for Monte 
Carlo - Data Comparison at HERA } in {\em Future Physics at HERA},
G.~Ingelman, A.~De Roeck, R.~Klanner (eds.), 
Proc. of the Workshop, DESY (1996), 611. \\
(see also http://www.desy.de/$\sim$h01rtc/hztool.html)

\bibitem{baryons} H1 Collaboration, C.~Adloff {\em et al.},
\Journal{\EJC}{6}{1999}{587}.

\bibitem{actw} 
L.~Alvero, J.~Collins, J.~Terron J.~Whitmore,
\Journal{\PRD}{59}{1999}{074022}; \\
 L.~Alvero, J.~Collins, J.~Whitmore: 
{\em Tests of Factorisation in Diffractive Charm Production and 
Double Pomeron Exchange}, hep-ph/9806340.

\bibitem{gammap} 
ZEUS Collaboration, J.~Breitweg {\em et al.}, 
\Journal{\EJC}{11}{1999}{35}; \\
H1 Collaboration, C.~Adloff {\em et al.}, 
\Journal{\PLB}{483}{2000}{36}.

\bibitem{incljets} H1 Collaboration, C.~Adloff {\em et al.}, 
\Journal{\EJC}{13}{2000}{415}.

\end{thebibliography}
\end{document}